\documentclass[11pt,a4paper]{article}
\pdfoutput=1
\usepackage{jcappub}
\usepackage{lineno}
\usepackage{epstopdf}
\usepackage{subfigure}
\usepackage{rotating}
\usepackage{float}

\usepackage[english]{babel}
\usepackage[utf8]{inputenc}

\graphicspath{{slike/},{slike2/},{slike3/}}

\newcommand{\be}{\begin{equation}}
\newcommand{\ee}{\end{equation}}

\newcommand{\lya}{Lyman-$\alpha$}
\newcommand{\lyb}{Lyman-$\beta$}
\newcommand{\Paa}{P_{\alpha\alpha}}
\newcommand{\Pbb}{P_{\beta\beta}}
\newcommand{\Pab}{P_{\alpha\beta}}
\newcommand{\Qab}{Q_{\alpha\beta}}

\newcommand{\Mpch}{\;\mathrm{Mpc\,h^{-1}}}
\newcommand{\hMpc}{\;\mathrm{h\,Mpc^{-1}}}
\newcommand{\skm}{\;\mathrm{s\,km^{-1}}}
\newcommand{\kms}{\;\mathrm{km\,s^{-1}}}
\newcommand{\K}{\;\mathrm{K}}

\newcommand{\Tref}{T_{\mathrm{ref}}}
\newcommand{\gref}{\gamma_{\mathrm{ref}}}

\title{The \lyb\ forest as a cosmic thermometer}

\author[a]{Vid Ir\v{s}i\v{c},}
\author[b,c]{Matteo Viel}

\affiliation[a]{The Abdus Salam International Centre for Theoretical
  Physics, Strada Costiera 11, I-34151 Trieste, Italy}
\affiliation[b]{INAF - Osservatorio Astronomico di Trieste, Via
  G. B. Tiepolo 11, I-34143 Trieste, Italy}
\affiliation[c]{INFN-National Institute for Nuclear Physics,
via Valerio 2, I-34127 Trieste, Italy}

\emailAdd{virsic@ictp.it}
\emailAdd{viel@oats.inaf.it}

\date{\today}

\abstract{ We present a comprehensive analysis of high resolution
  hydrodynamic simulations in terms of \lya\ and \lyb\ one dimensional
  flux power spectra ($\Paa$ and $\Pbb$).  In particular, we focus on
  the behaviour that the flux auto-power spectra and cross-power
  spectra ($\Pab$) display when the intergalactic medium (IGM) thermal
  history is changed in a range of values that bracket a reference
  model, while cosmological parameters are kept fixed to best fit the
  cosmic microwave background data. We present empirical fits that
  describe at the sub-percent level the dependence of the power
  spectra on the thermal parameters. At the largest scales, the power
  spectra show a constant bias between each other that is set by the
  parameters describing the IGM thermal state. The cross-power
  spectrum has an oscillatory pattern and crosses zero at a scale
  which depends on $T_0$, the IGM temperature at the mean density, for
  reasonable values of the power-law index $\gamma$ of the IGM
  temperature-density relation ($T=T_0(1+\delta)^{\gamma-1}$). By
  performing a Fisher matrix analysis, we find that the power spectrum
  $\Pbb$ is more sensitive to the thermal history than $\Paa$ alone,
  due to the fact that it probes denser regions than \lya\. When we combine the
  power and cross spectra the constraints on $\gamma$ can be improved
  by a factor $\sim 4$, while the constraints on $T_0$ improve by a
  factor of $\sim 2$. We address the role of signal-to-noise and
  resolution by mocking realistic observations and we conclude that
  the framework presented in this work can significantly improve the
  knowledge of the IGM thermal state, which will in turn guarantee
  better constraints on IGM-derived cosmological parameters.}

\begin{document}

\maketitle

\section{Introduction}

The last decade has witnessed an enormous progress in the cosmological
investigation of the intergalactic medium (IGM) as probed by the
\lya\ forest (see \cite{rauch98,meiksin09} for reviews).  In
particular, the \lya\ forest is now a viable cosmological observable
that can help in putting constraints on cosmological parameters and/or
deviations from a standard scenario based on cold dark matter and a
cosmological constant.  Several statistics of the transmitted flux can
be considered. For example, the one-dimensional (1D) \lya\ flux power
spectrum
\cite{mcdonald00,zaldarriaga01,croft02,k04,vhs04,mcdonald06,palanque13},
as derived from both high and low-resolution data, is the quantity
that has been widely used in order to place constraints on the
amplitude and slope of the linear matter power spectrum in a unique
range of redshifts and scales \cite{croft02,mcdonald05,vhs04}.  This
has in turn allowed to put tight limits on fundamental
quantities like inflationary parameters, cold dark matter coldness and
neutrino masses \cite{seljak06,viel05}, especially when these data are
combined with large scale information like the cosmic microwave
background and galaxy clustering. Basically, any physical effect, like
for example modified gravity, happening at scales of $\sim 1-100$
comoving Mpc$/h$ will impact dramatically on the \lya\ forest
properties as probed by the 1D flux power.  More recently, the very
large number of quasar spectra of the SDSS-III/BOSS (Sloan Digital
Sky Survey-III/Baryon Oscillation Spectroscopic Survey) collaboration
using a state-of-the-art analysis, has allowed to discover
Baryonic Acoustic Oscillations (BAOs) in the transmitted \lya\ flux at
$z\sim 2.3$ \cite{busca13,slosar13}: this spectacular
confirmation of the cosmological nature of the \lya\ forest was mainly
made possible by exploiting the flux correlations in three dimensions
out to large scales.

What is usually not fully appreciated is that final cosmologically
relevant numbers are obtained by running a relatively sophisticated
pipeline that marginalizes over the many nuisance and/or astrophysical
parameters affecting the investigated IGM observable.  The parameters
describing the physics of the IGM are in principal many, since the
physics is rich and affected by astrophysical properties of the
surrounding galaxies (e.g. galactic feedback in the form of galactic
winds or Active Galactic Nuclei or Blazar heating
\cite{mcdonald05b,vielschaye,puchwein12}), or by the presence of
strong systems that are usually associated with dense environments which
are difficult to model \cite{carswell,mcdonald05b}, or reionization
and complex radiative transfer effects impacting on the low density IGM \cite{mcquinn}.

However, there are more obvious parameters that are also constrained
independently by different techniques: those governing the so-called
``equation-of-state'' of the IGM. This relation is often assumed to be
adiabatic in form, connecting the temperature ($T$) to the baryon
density $\rho_{\rm b}$:
\be T = T_0
\left(\frac{\rho_b}{\rho_{\mathrm{mean}}}\right)^{\gamma-1} = T_0
\Delta_b^{\gamma-1}, 
\label{eq:T-rho-adiabatic}
\ee where
$\rho_{\mathrm{mean}}$ is critical density of the Universe at a given
redshift and $\Delta_b$ is the baryon density contrast. The parameters
are $T_0$ and $\gamma$, governing the mean
temperature of the IGM and the slope of the
$T-\rho_b$ relation, respectively \cite{huignedin97,huihaiman97}.  Note
that these parameters have been constrained in a series of works by
using: line-fitting procedures
\cite{schaye00,ricotti00,rudie12,bolton14}, wavelet analysis
\cite{theuns00,theuns02,lidz10,garzilli}, flux probability
distribution function \cite{boltonpdf,KGpdf14}, flux power spectrum
\cite{mcdonald00,vh06}, and curvature of the transmitted flux
\cite{becker11,boera14}.  All these different methods provide
constraints (at different redshifts and probing slightly different
density regimes) with still relatively large error bars.

While degeneracies are intrinsically present in the \lya\ forest
analysis, one can hope that with better constraints on the parameters
of the IGM, the strength of the degeneracy can be lifted. 
To break up this degeneracy several studies have suggested to use
either higher order statistics of the \lya\ forest (such as
bispectrum, flux pdf, wavelet) (see e.g. \cite{lidz14}) or higher
order neutral hydrogen transitions (such as \lyb\ or Lyman-$\gamma$
transitions), or to exploit the (possibly different) redshift
dependence of cosmological and astrophysical effects by using a wider
redshift range (e.g. like in BOSS as analyzed by \cite{mcdonald05}).

Using only \lya\ forest to determine the IGM parameters is difficult
since the \lya\ cross section is large enough to produce saturated
absorption fast. The \lyb\ forest presents an opportunity to measure the same
large-scale structure of \lya\ but with a different sensitivity
to the optical depth of the photons travelling through the IGM. In other
terms, \lyb\ forest is an additional measurement of the IGM and large
scale structure but with different bias parameters. However, since
the bias relationship between the observed flux and the optical depth
is highly nonlinear, the relationship between the $\Paa$ and $\Pbb$
is not trivial.

So far the gain of using the \lyb\ forest has been estimated
using dark matter N-body simulations \cite{lidzlyb}. On the data
side, a recent investigation of \lyb\ flux power spectrum, and the
cross spectrum using BOSS data has been presented in
\cite{irsic13}. Even though these findings provide convincing
indications that using the \lyb\ forest statistics should alleviate
the tensions between IGM parameters and cosmology, a more complete and
precise assessment using hydrodynamical simulation is deemed
necessary. In this respect, the present work is an improvement of the
already existing analysis with the important new ingredient of probing
also the cross-spectrum between \lya\ and \lyb.

The paper is structured as follows: in Section 2 we describe the
hydrodynamic simulations used; in Section 3 we discuss how we model the
\lyb\ absorption; Section 4 contains our results in terms of auto and
cross spectra together with a quantitative analysis of their dependence
on the thermal state; Section 5 presents the Fisher Matrix analysis 
of the power spectra; Section 6 summarizes our findings.

\section{Hydrodynamic Simulations and Spectra Analysis}
The grid of simulations used in the present analysis is performed by
using the Tree-Particle Mesh (PM) Smoothed Particle Hydrodynamics
(SPH) code {\sc GADGET-III} \cite{gadget}. Star formation is followed
by using a simplified procedure (commonly denoted {\sc
  QUICKLY$\alpha$}), employed for the first time in \cite{vhs04},
which turns gas particles with overdensity larger than
1000 and temperature below $10^5$ K into stars. This procedure has been shown to
provide sub-percent level accuracy in terms of \lya\ statistics when
compared to the more refined effective model for the inter-stellar
medium \cite{vhs04}.  The cooling routines have been slightly modified
in order to achieve a given thermal history at the desired redshift as
done recently in \cite{vielwdm}, where the grid of thermal histories
explored in the present work is also presented. Thereby, we do not
make an a-posteriori scaling of the temperature density relation but
our thermal histories have been obtained consistently within the
SPH simulations. Our reference thermal history is tailored to match
the recent curvature-based measurements of $T_0$ presented in
\cite{becker11}.

The simulated cosmological periodic volume has a linear size of
$60\Mpch$, $512^3$ gas particles and $512^3$ dark matter particles
(the mass per gas particle is $m_{\rm p}=2\times10^7\, M_{\odot}/h$).
The cosmological parameters, kept fixed, have the following values:
$\Omega_{\rm m} = 0.27$, $\Omega_{\rm b} = 0.0458$, $\Omega_\Lambda =
0.72$, $n_{\rm s}=0.968$, $H_0=70.2$ km/s/Mpc, $\sigma_8=0.816$ which
are in agreement both with WMAP-9yr and Planck \cite{wmap9yr,planck}.

We consider the snapshots at one redshift ($z=3$) and for different
thermal histories with colder or hotter IGM ($T_0$) and lower or
higher $\gamma$, with respect the reference model. The different thermal histories are constructed by
modifying the fiducial simulation He II photo-heating rate,
$\epsilon^{\mathrm{fid}}_{{\mathrm{HeII}}}$, such that
$\epsilon_{\mathrm{HeII}} = \alpha
\Delta_b^\beta\epsilon^{\mathrm{fid}}_{{\mathrm{HeII}}}$. The
procedure is described in more detail in \cite{2008MNRAS.386.1131B}.

\begin{figure}
  \centering
  \includegraphics[width=1.0\linewidth]{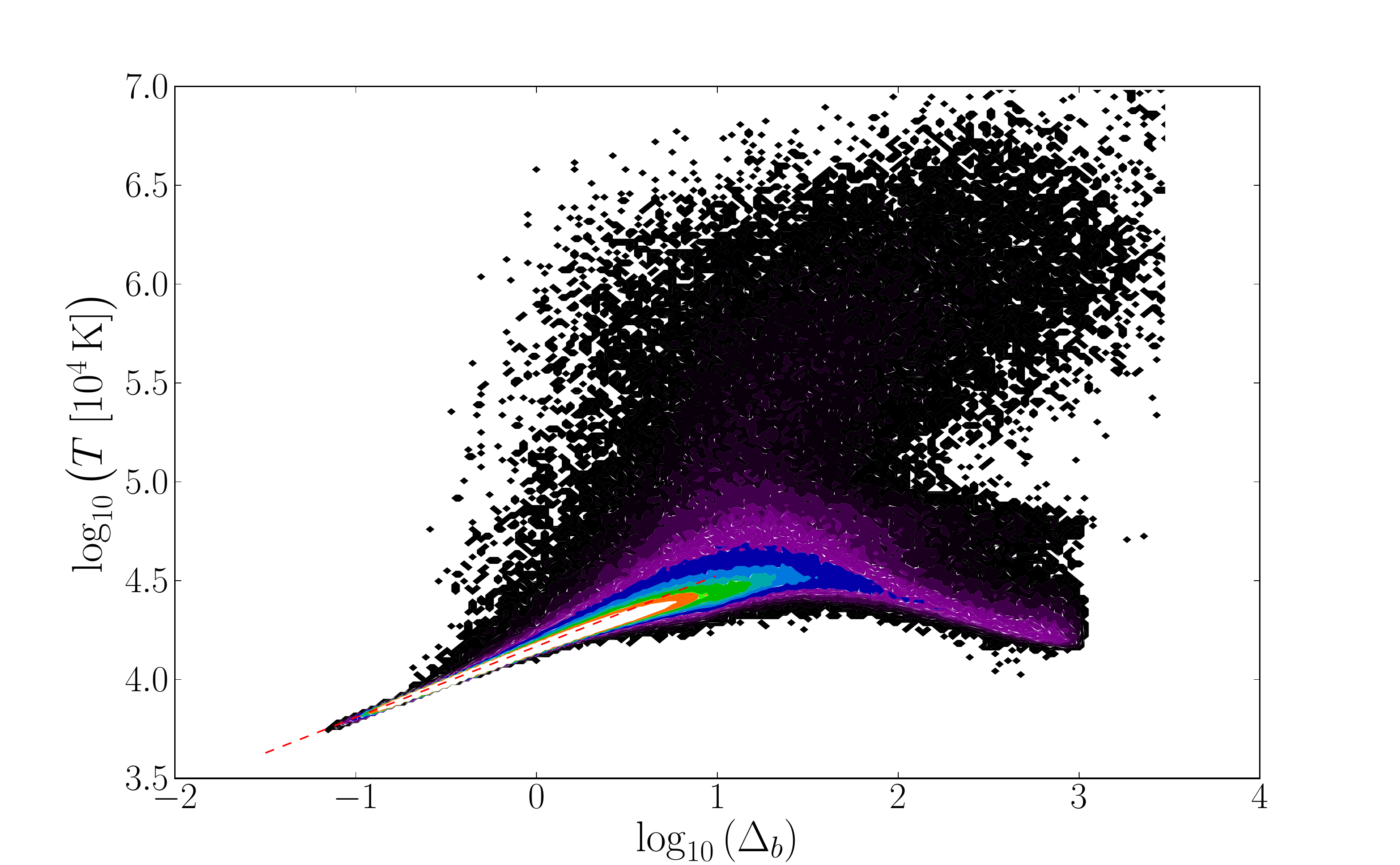}
  \caption{
    The figure shows the contours of the scatter plot of the $T-\rho$
    relation for the reference model (REF). Different colours
    represent increasing number density of points. Dashed red line
    shows the power law approximation used to determine the
    characterize the temperature-density relation. The part of the
    scatter plot used to determine the power law parameters is
    only in the ranges of $-1<\log_{10}\left(\Delta_b\right)<0$
    and $\log_{10}\left(T_4\right) < 5$ (where $T_4 = T/10^4\;K$).
  }
    \label{Fig:Trho}
\end{figure}

The values of $T_0$ and $\gamma$ are measured from the scatter
$T-\rho_b$ plot (Fig. \ref{Fig:Trho}) 
using only the points with low temperature ($T_0 <
10^5\K$). Moreover, all the points that are too
overdense ($\Delta_b > 1$) are discarded to take into account
only the region of the $T-\rho_b$ space where the relation can be
modeled with a simple adiabatic form
(Eq. \ref{eq:T-rho-adiabatic}). Additionally, the points too
underdense ($\Delta_b < 0.1$) are not used as well, because our
simulations do not have enough resolution to properly model the density
field in voids. The remaining points are used in a
linear regression model to determine the parameters of the
temperature-density relation (in a log-log plane).

The different simulations used are summarized in \ref{tb1}.

\begin{table}[H]
\centering
\begin{tabular}{|c|c|c|c|c|}
\hline
Model & thermal history & $L_{{\rm box}}\;[\Mpch]$ & $N_g$ & $N_{\rm dm}$ \\
\hline\hline
L60 & REF+& $60$ & $512^3$ & $512^3$ \\ \hline
L30 & REF & $30$ & $512^3$ & $512^3$ \\ \hline
L15 & REF & $15$ & $512^3$ & $512^3$ \\ \hline
L15s & REF & $15$ & $128^3$ & $128^3$ \\ \hline
\end{tabular}
{
  \caption{Simulations used in this work: simulations with different
    resolutions were computed at the reference (REF) thermal history
    (see table \ref{tb2}). Thermal
    history REF+ refers to the reference (REF) thermal history as well as all the
    others described in the table \ref{tb2}. }
}

\label{tb1}
\end{table}

\begin{table}[H]
\centering
\begin{tabular}{|c|c|c|c|c|c|}
\hline
Model & descr. & $T_0\;[\mathrm{K}]$ & $\gamma$ \\
\hline\hline
REF & Ref & $14709$ & $1.359$ \\ \hline
COLD & Cold & $9820$ & $1.358$ \\ \hline
HOT & Hot & $19593$ & $1.360$ \\ \hline
LG & Low$\gamma$ & $15036$ & $1.032$ \\ \hline
HG & High$\gamma$ & $14499$ & $1.580$ \\ \hline
HHG & Hot + High$\gamma$ & $19346$ & $1.580$ \\ \hline
HLG & Hot + Low$\gamma$ & $19980$ & $1.033$ \\ \hline
CHG & Cold + High$\gamma$ & $9661$ & $1.579$ \\ \hline
CLG & Cold + Low$\gamma$ & $10068$ & $1.029$ \\ \hline

\end{tabular}
{
  \caption{Simulations used in this work: thermal parameters that
    parameterize the $T-\rho_b$ relation are reported.}
}

\label{tb2}
\end{table}

We extract a number of randomly determined lines-of-sight (LOSs)
through the simulation box at a given redshift and interpolate the gas
properties (density, temperature and velocity) along them. This
enables us to determine the optical depth ($\tau$) for each pixel, and
in turn the observed flux ($F$). The interpolation scheme adopted in
this paper is described in more detail in \cite{theuns98}. The number
of pixels along each LOS is chosen to be $N_{\rm o} = 2048$. The
results are robust to this choice (up to a high value for the
wavenumber $k$ of the recovered power spectrum, which is enough for
the purposes of this paper). The number of LOSs is $N_{\rm l} = 5000$
and we have tested that the numbers above are high enough for the
statistical estimators to converge.

Once the optical depths due to both \lya\ and \lyb\ absorption along
all LOSs have been determined, we first re-scale the \lya\ optical
depth so that the mean observed flux matches observations at a given
redshift. In practice, we rely on the mean flux level measurement
inferred from high-resolution spectra presented in \cite{kim07} which
is usually casted in the form of an effective optical depth: $\tau_{\rm
  eff}=0.0023\times(1+z)^{3.65}$. The re-scaling procedure can be interpreted as
scaling of the mean Ultraviolet (UV) background flux and we also
re-scale optical depth for \lyb\ absorption by the same amount. We are
allowed this freedom because the relation between mean observed flux
and mean UV flux is degenerate and the first is more stringently
constrained by observations \cite{peeples10a,peeples10b}.  We perform
this step regardless of the thermal history used in the simulations,
but we caution the reader that the mean observed flux depends also on
$T_0$ and $\gamma$. The rescaling typically changes the optical depths
by an amount which is of the order of $10-20$\%.

The optical depth skewers (in both \lya\ and \lyb) are then converted
into observed flux fluctuations. Those are transformed in the
wave-vector ($k$) space using Fast Fourier transformation (FFT)
techniques using the freely available {\sc FFTW} library. In the end,
the coefficients of the Fourier transform are squared and averaged
over all the LOSs to get an estimate on the (auto) power spectrum. The
cross power spectrum is obtained by multiplying coefficients of
\lya\ flux with those from \lyb. Since this analysis is only concerned
with the recovering of the power spectra at a given redshift and does
not model the redshift evolution along the LOS of optical
depth,  the recovered cross power spectrum will not have any imaginary
part contribution, as suggested by \cite{irsic13}.

To estimate the error-bars of power spectra we adopt the bootstrap
technique. By using randomly determined weighted sum of the Fourier
coefficient contribution from each LOS we average over the weights
using $N_{\rm b}=1000$ bootstrap samples. We show that such a number
of samples is enough to achieve converged results for the whole
covariance matrix of the $k$-bins of each of the power
spectra. However, this technique lacks the ability to add covariance
between the estimated power spectra, which is something that is not
addressed in the present paper.

\section{Modelling \lyb\ absorption}
To model the \lya\ forest absorption we adopt a standard framework
which allows us to extract randomly selected skewers through the
simulation box at a given redshift. For each pixel along each
LOS we interpolate gas related physical quantities and sum
over all the contributing gas particles that affect the result within
the window function used, thereby relying on a Smoothed Particle
Hydrodynamics (SPH) scheme to obtain physical quantities at each
pixel. The formalism used is described in more detail in
\cite{theuns98}. We consider both the full Voigt profile model of each
absorption feature as well as correct for peculiar velocities of the
gas along the line-of-sight. The same technique has been used multiple
times in many different analysis investigating the \lya\ forest flux
statistics like flux probability distribution function, flux power
and bispectrum etc. (e.g. \cite{vhs04,vhsbisp,boltonpdf}).

However, in order to take \lyb\ absorption into account as well, we
extend the model above. We assume that the only difference between
\lya\ and \lyb\ absorption comes from different absorption cross
sections (main effect) and different natural widths of Lorentz
profiles of those two transitions (secondary effect). This implies
that the \lya\ and \lyb\ optical depths are related as:

\be
\tau_\alpha(x) = \sigma_\alpha \tau(x,\Gamma_\alpha), \qquad
\tau_\beta(x) = \sigma_\beta \tau(x,\Gamma_\beta), 
\ee 

where the function $\tau(x,\Gamma)$ is a general function of optical
depth describing all the physics due to fluctuating density field, gas
properties and modelling of the line profiles. By changing the
oscillator strength $f$, rest-frame wavelength of the transition
($\lambda$) and the natural width $\Gamma$ we can use the same
formalism for \lyb\ and for \lya. 

The values of the atomic physics constants used in our calculations
were (as extracted from {\sc VPFIT} {\footnote{Version 10.0 by R.F. Carswell
    and J.K. Webb, http://www.ast.cam.ac.uk/rfc/vpfit.html}}:
HI-Ly$\alpha$ ($\lambda_\alpha = 1215.6701$\AA, $f_\alpha = 0.4164$,
$\Gamma_\alpha = 6.265\,\times\,10^8\;\mathrm{s^{-1}}$) and
HI-Ly$\beta$ ($\lambda_\beta = 1025.7223$\AA, $f_\beta = 0.07912 $,
$\Gamma_\beta = 1.897\,\times\,10^8\;\mathrm{s^{-1}}$).

\section{Results}
\begin{figure}
  \centering
  \includegraphics[width=1.0\linewidth]{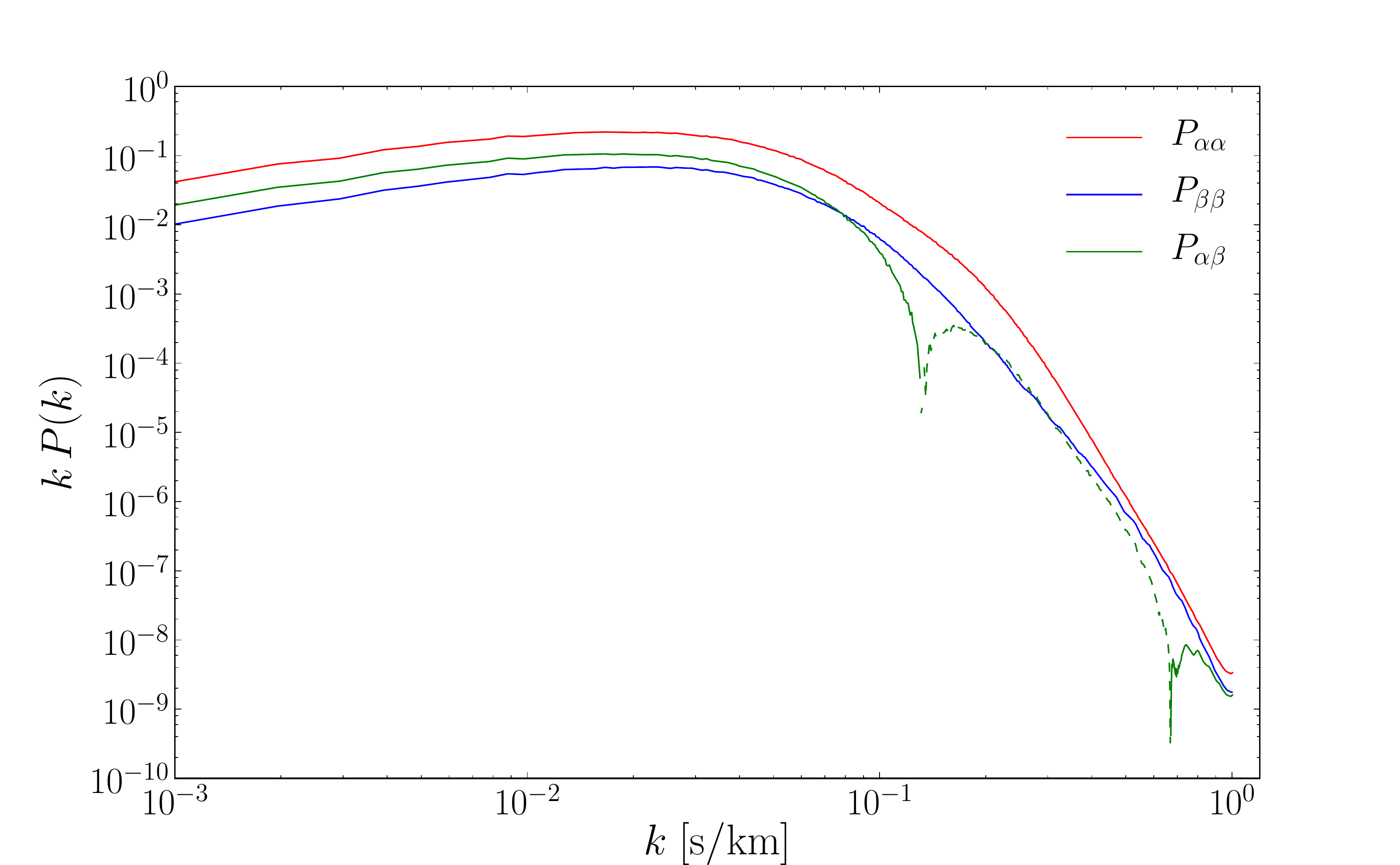}
  \caption{Power spectra extracted from the hydrodynamic
    simulations. The simulation box is $60\Mpch$ in size with $2\times
    512^3$ dark matter and gas particles, for the reference (REF)
    thermal history simulation. The power spectra shown are for $\Paa$
    (red), $\Pbb$ (blue) and $\Pab$ (green). Dashed green line shows
    the absolute value of the cross power in a $k$-range where it is
    negative. On large scales power spectra are linearly biased
    towards each other. On small scales we see the effects of
    non-linearity of the $F-\tau$ relation, where we can see both
    $\Paa$ and $\Pbb$ curves falling off and getting closer
    together. An important feature is also the zero crossing of the
    cross power spectrum, $\Pab$, at around $k\sim 0.12\skm$ (and
    again at $k\sim 0.65\skm$) indicating an oscillatory
    pattern. No resolution correction was added to any three of the flux
    power spectra in this plot (see Sec. \ref{sec:res-corr}).}
     \label{Fig:1}
\end{figure}

In this section we present our main results.  The power spectra
obtained from the simulation at redshift $z=3$ are shown in
Fig. \ref{Fig:1}. Although it is clear that on large scales
($k<0.01\skm$) the three power spectra are in a regime of a constant
bias with respect to each other, the small scales highlight the trends
of non-linearity due to the $F-\tau$ transformation.

If the contribution of the different Lorentzian natural widths of the
respective lines are neglected, then the relation between $\tau_\alpha$
and $\tau_\beta$ is linear, whereas the relation between the observed
fluxes $F_\alpha$ and $F_\beta$ is not linear on small scales.  The
observed trend of $\Pbb$ and $\Paa$ getting closer with higher $k$ is
a result also obtained from N-body simulations using dark matter only
\cite{lidzlyb} and can be interpreted as a result of the non
linear nature of the transformation. On sufficiently small scales, the
power spectra $\Paa$ and $\Pbb$ grow in a very similar way
indicating that both power spectra are tracing the same underlying line
structure. Eventually, this behaviour breaks down at even smaller
scales where the effect of different natural broadening of the lines
comes into play.

\begin{figure}
  \centering
  \includegraphics[width=1.0\linewidth]{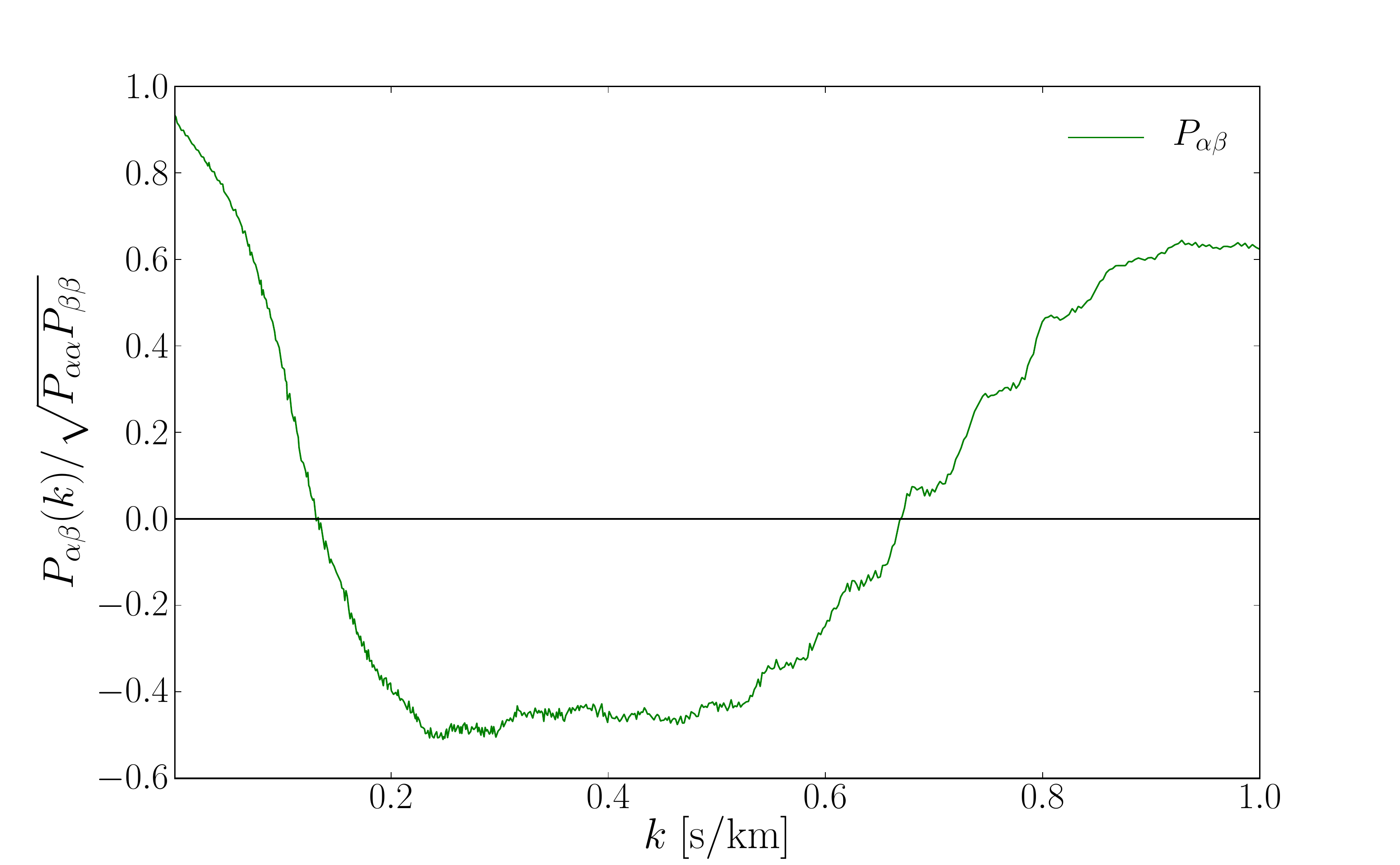}
  \caption{Cross power spectrum $\Pab$ divided by the geometric mean
    of both auto power spectra ($\Paa$ and $\Pbb$) for the reference
    (REF) simulation. At large scales, the cross correlation
    coefficient is slightly below one, while on
    smaller scales it falls off as expected. Non-linearity of the
    transformation flux-optical depth makes it (roughly) constant at
    intermediate scales (in a way which depends on the thermal
    history). However, independently of the thermal history results,
    the cross power spectrum grows again on very small scales where it
    is believed to trace the structure of each individual
    absorption line. From large to small scales $\Pab$ thus makes
    roughly one oscillation of what seems to be quite a substantial
    fluctuation. Note also the smaller oscillatory pattern at the
    smallest scales which is linked to the correlations between
    individual lines. }
  \label{Fig:2}
\end{figure}

Another very interesting feature to notice is that at small scales ($k
\sim 0.1\skm$, corresponding to roughly $k \sim 10\hMpc$) the cross
power spectrum ($\Pab$) crosses zero. In fact, Fig. \ref{Fig:2} shows
that the cross power spectrum exhibits an oscillation pattern. The
normalized cross power spectrum, as plotted in Fig. \ref{Fig:2},
measures the cross-correlation coefficient which is defined as:
\footnote{This is exactly true in the absence of the imaginary part
  of the cross power spectrum $\Qab$.}  
\be r_{\alpha\beta}^2 = \left[
  \frac{\Pab^2 + \Qab^2}{\Paa \Pbb} \right].
\label{eq:rab}
\ee

From the definition  above, we see that the cross-correlation coefficient
takes strictly positive values. However, in our estimate we choose to preserve
the sign of the $\Pab$ to stress the features of the oscillatory
pattern. Since both $\Paa$ and $\Pbb$ are positive in our measurements
(by definition) the transformation of the plot in Fig. \ref{Fig:2} to
$r_{\alpha\beta}$ is easily achieved.

Using the hydrodynamic simulations we confirm the findings that the
cross-correlation coefficient goes towards $1$ at large scales and
falls off at smaller scales, as was measured in BOSS DR9 by
\cite{irsic13}. 
However, we also note
that on small scales the cross-correlation coefficient does not go to
zero, but rather exhibits an oscillation.

While the details of the oscillations in $\Pab$ vary among
simulations with different thermal histories, the overall trend
remains the same. Since we also note that the oscillation is only weakly
damped, we conclude that its nature is due to the non-linearity of the
transformation between the density field and the observed flux, rather
than carrying physical meaning. However, we caution the reader
that the results of this paper are not conclusive on the point of the
nature of $\Pab$ oscillatory pattern. Further investigation is
required to confirm the interpretation given. 

Another oscillatory pattern can be seen for $\Pab$ 
in Fig. \ref{Fig:2} albeit with a much smaller amplitude and scale,
showing up at the smallest of scales ($k > 0.3-0.5\skm$) even in
$\Pbb$. However, this paper is focused on larger scales ($k <
0.1\skm$) and trends could be due to the cross correlation of
individual lines that is not the subject of the present investigation.

\subsection{The resolution and convergence tests}
\label{sec:res-corr}

To test the stability and convergence due to simulation resolution on
small scales we have performed additional flux power spectra analysis
on two higher resolution simulations: $30\Mpch$ box size with $2\times 512^3$
gas and dark matter particles (L30) and $15\Mpch$ box size with the same
number of particles (L15). The details of the simulations are summarized in
table \ref{tb1}.  The higher resolution simulations (L30 and L15) have
the same cosmological parameters as used in the L60 simulation
suite. However, only the reference (REF) thermal history was considered
for the reasons that will be described shortly.

\begin{figure}
  \centering
  \includegraphics[width=1.0\linewidth]{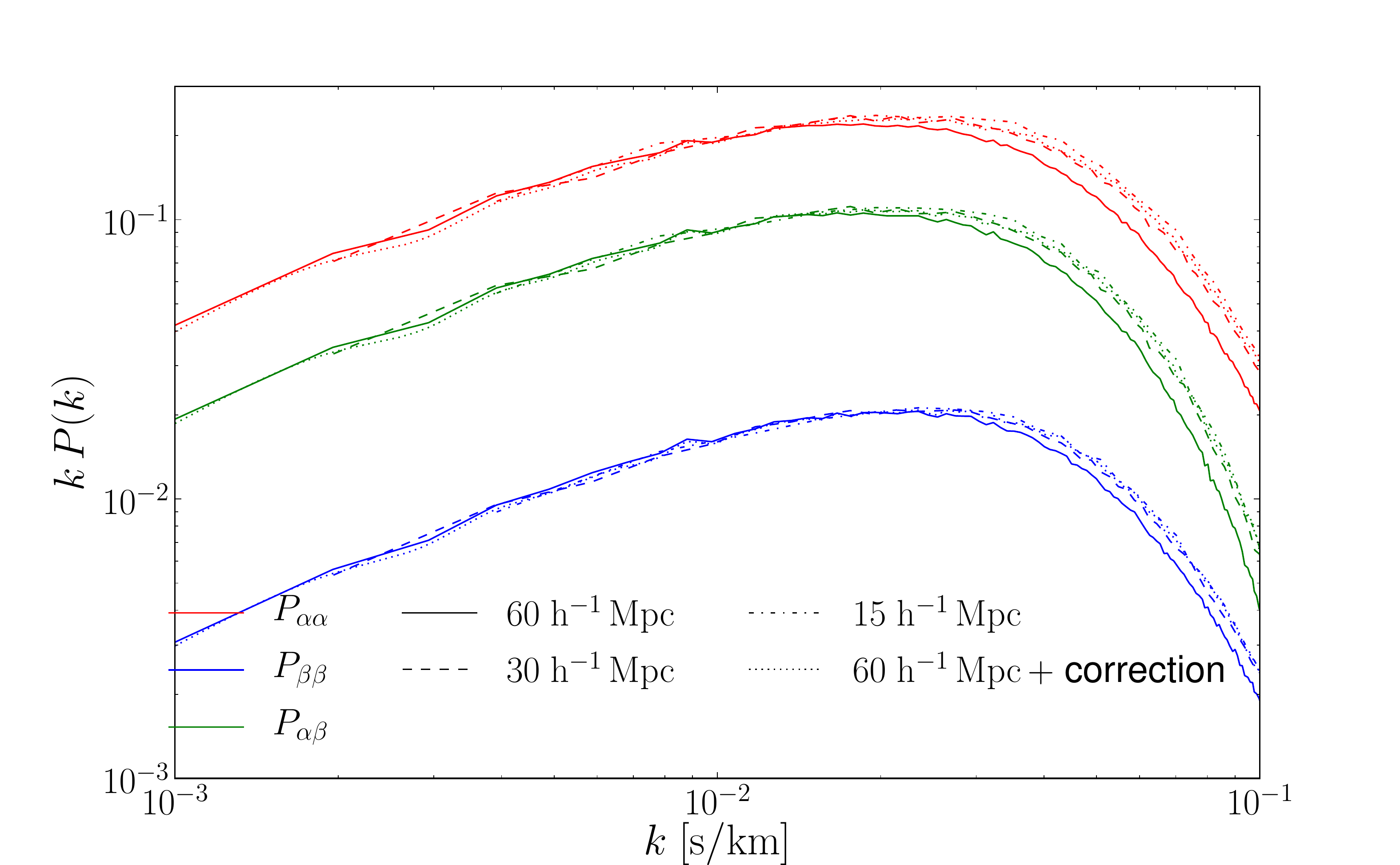}
  \caption{ 
    The figure shows the resulting flux power spectra for different
    resolutions of the simulations: $\Paa$ (red),
    $\Pbb$ (blue) and the cross-power $\Pab$ (green). The power
    spectra on this plot are shifted for clarity reasons.
    Different
    line styles  represent different simulations with L60 (full line),
    L30 (dashed line), L15 (dot-dashed line) (see Table \ref{tb1} for
    details). All simulations were computed using the same (reference)
    thermal history and same cosmological parameters. With dotted line
    we show the result of corrected flux power spectrum (see
    text).
   }
  \label{Fig:3a}
\end{figure}

Figure \ref{Fig:3a} shows the result of changing the resolution of the
entire simulation, by keeping the number of both gas and dark matter
particles fixed and decreasing the size of the simulated box. As has
been noted by other similar resolution tests (\cite{mcdonald03,lukic,borde}) the effect of increasing resolution is the increase
of power on small scales, while the large scales remain mostly
intact. However, if the resolution would have been decreased further
by increasing the size of the box beyond $60\Mpch$, having fixed the number of particles, an effect on large scales would be seen as well. It
is also clear that the convergence of power is reached somewhere
between the resolution of L30 and L15, since the curves do not differ
from one another. Thus, we can assume that on small scales ($k >
0.02\skm$) the resolution of L15 is good enough to capture the physics
of the \lya\ (and \lyb) forest.

In principle one could correct for this effect using suit of 3
simulations: large box size (but lower resolution) one, and two small
box size simulations. Of the two small box size simulations one would
have the same number of particles as the large box size simulation
(and thus higher resolution) and the other would have the same
resolution as the original large box size simulation. Such a
correction was first proposed in \cite{mcdonald03} and is described in
more detail therein. 

\begin{figure}
  \centering
  \includegraphics[width=1.0\linewidth]{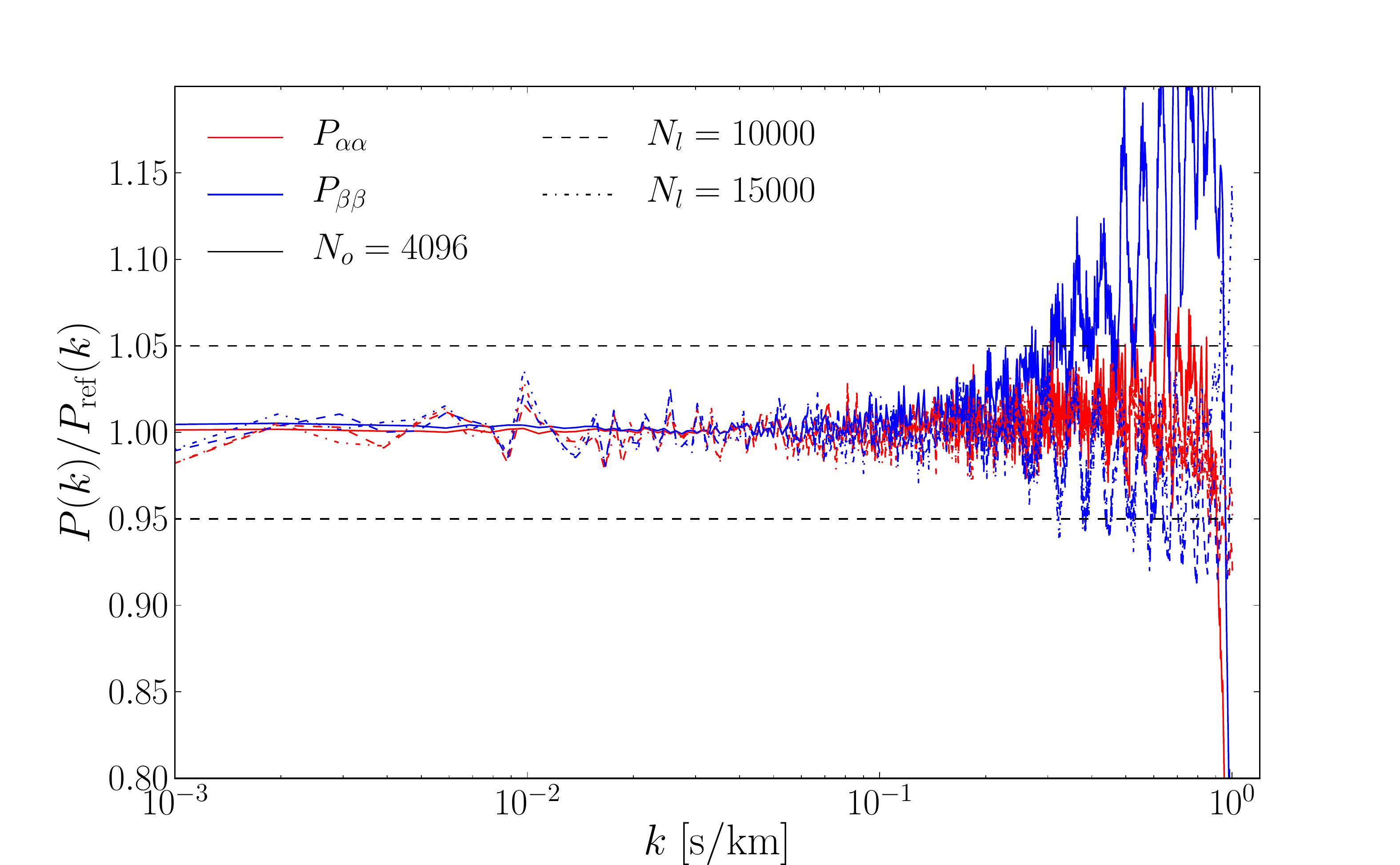}
  \caption{ This plot shows the dependence of the power spectra -
    $\Paa$ (red) and $\Pbb$ (blue) - on the parameters of the analysis
    pipeline (number of pixels and number of LOSs ($N_{\rm o}$ and
    $N_{\rm l}$, respectively). 
    On y-axis we plot the ratio between
    the power spectrum with changed parameters towards the power
    spectrum using the reference values (REF). The full line shows how the
    power spectra change when increasing the number of bins along each
    line of sight at which we evaluated interpolated quantities of the
    SPH simulation. The dashed line shows the changes to the power
    spectra when increasing the number of LOSs in our analysis.
    We can see that at large scales  differences are within a few percent level, while they grow quickly on small
    scales, exceeding the 5\% in $\Pbb$ at scales larger than $k\sim
    0.3\skm$, while the results for $\Paa$ are more stable. 
  }
  \label{Fig:3}
\end{figure}

The result of the corrected flux power spectrum for the resolution is
shown in Fig. \ref{Fig:3a} with dotted line. We have only corrected
for large-scale normalization and intermediate scales, but not for the
low scales where L60 does not sample the power correctly anymore,
since our focus in this work is on larger scales. The correction works
reasonably well.

It is believed that at first order this kind of correction is
independent of the thermal history of the gas in simulations, thus
only one (reference) thermal history case needs to be explored to
calibrate the correction relations. This might not be
entirely true at very high precision on small scales, but we have
not explored this issue further. 

However, since our final tests and results are obtained on the ratios
of flux power spectra between different thermal history simulations
and the reference one, any such (thermal history independent)
correction for resolution would cancel out completely. This is one of
the advantages in working with ratios and was one of the reasons we
have chosen such an approach. 

We tested the sensitivity of the power spectra to the changes of
parameters of the analysis process, namely the number of LOSs
($N_{\rm l}$), over which the flux power spectrum is averaged and the number
of pixel bins along each line of sight $N_{\rm o}$. Figure \ref{Fig:3} shows
that on larger scales ($k < 0.1\skm$) the power spectra have converged
and the remaining differences are below a few percent level. At small
scales, however, the differences grow significantly.

Increasing the resolution along each LOS allows to widen the range
at which the power spectra can be measured. As it can be seen in
Fig. \ref{Fig:3}, a larger  number of pixel bins along the
LOS leads to higher values of $\Pbb$ at the very small scales
($k > 0.3\skm$), while $\Paa$ remains largely unchanged. This indicates
that at those scales $\Paa$ and $\Pbb$ grow even further together than
estimated from Fig. \ref{Fig:1}, further reassuring the point that on
those scales the power spectrum traces correlations among individual
lines and that the power spectra should be very similar in shape.

\subsection{Auto power spectra dependence on the IGM thermal state}
We now quantify the dependence of the power spectra $\Paa$, $\Pbb$ on
the parameters of the equation of state ($T_0,\gamma$), governing the
thermal history of the universe.

To address this issue we analyze the models of Table \ref{tb1}. All
the power spectra retrieved from simulations show a general tendency:
a change in the thermal history (either temperature or the adiabatic
index) changes the scale of an exponential cut-off on the smaller
scales, while modifications at the large scales are less
pronounced. Of course, the cut-off is not exactly exponential and
there are many features introduced in the slope. Also note that at
least on small scales there is a visible degeneracy between increasing
the temperature of the IGM and lowering the adiabatic index, with
different power spectra being sensitive to this change in a different
way.

\begin{figure}
  \centering
  \includegraphics[width=1.0\linewidth]{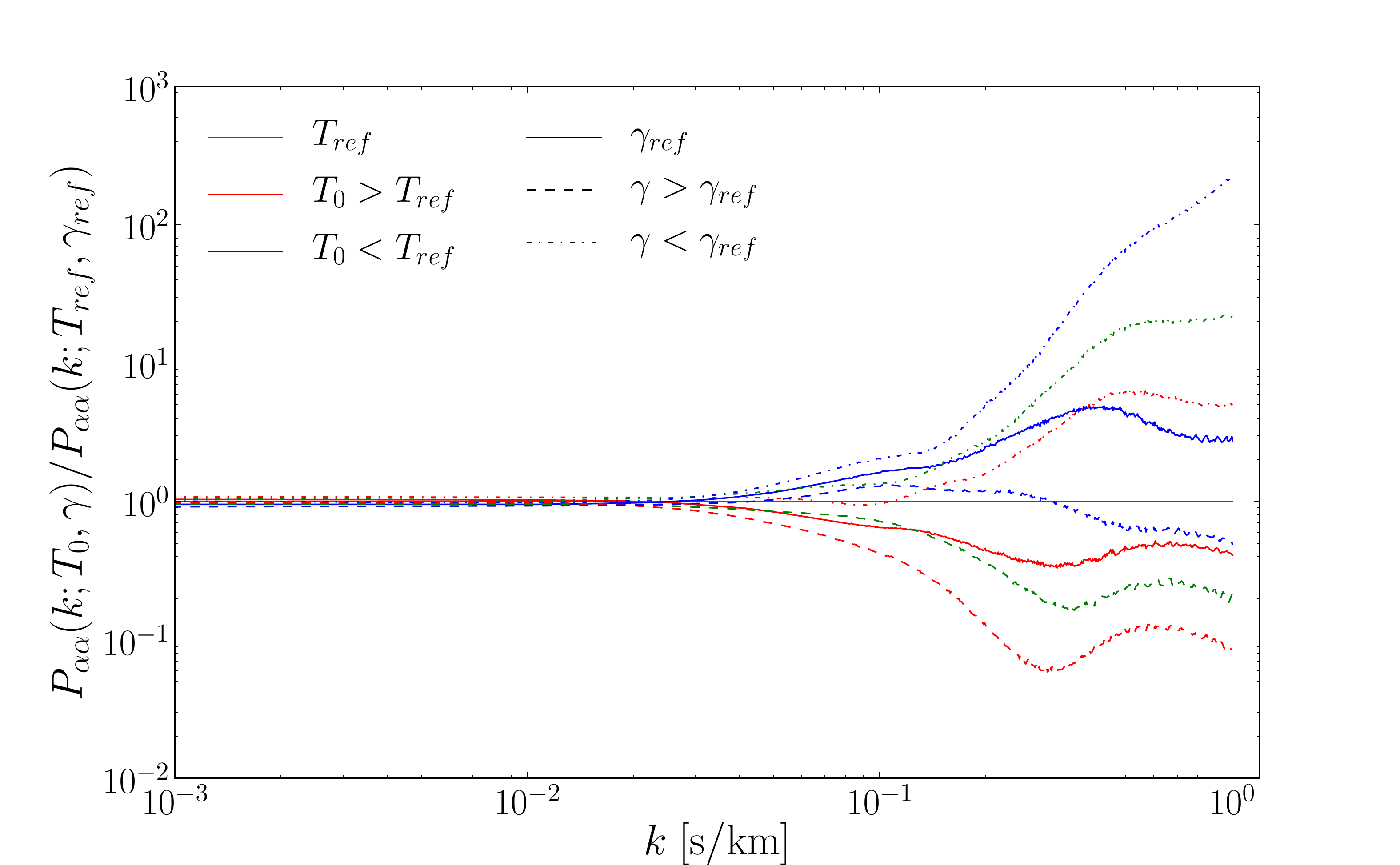}
  \caption{ \lya\ power spectrum for simulations with different
    thermal histories. Three different colours correspond to three
    different (mean density) temperatures of the gas ($T_0$) with
    green (models with $T_0 = \Tref$), red ($T_0 > \Tref$) and blue
    ($T_0 < \Tref$).  Different line styles correspond to different
    parameter of the slope of the $T-\rho$ relation ($\gamma$) with
    full line (models with $\gamma = \gref$), dashed line ($\gamma >
    \gref$) and dash-dotted line ($\gamma < \gref$). All the models
    shown can be found in Table \ref{tb1}.  $\Paa$ is
    very insensitive to changes in thermal history at large scales,
    while on small scales we note the recurring behaviour of larger
    $\gamma$ or hotter gas producing deficit of power on small
    scales. However, the difference between models is not very
    substantial which implies a relatively poor estimations of $T_0$ and
    $\gamma$ parameters when cosmological \lya\ only analysis is performed. }
  \label{Fig:4}
\end{figure}

The \lya\ power spectrum, shown in Fig. \ref{Fig:4}, offers the
general indications summed up in the previous paragraphs. While
the (relative) changes due to different thermal histories are small on
large scales ($k < 0.02\skm$), they are much more pronounced on small
scales. Different colours on the plot show different temperatures of
the IGM (green for reference temperature of $T_{\rm ref} =
14709\;\mathrm{K}$, blue for colder and red for hotter IGM) and
different line styles represent different values of the adiabatic
index (full line for reference value of $\gamma_{\rm ref} = 1.359 $,
dashed line for higher values and dash-dotted for lower values of the
adiabatic index). As noticed there is a slight degeneracy between varying
$T_0$ or $\gamma$, separately, e.g. a hot IGM with adiabatic index
closer to $1$ (red dash-dotted line) would produce similar results as
a cold IGM with reference adiabatic index (blue line). However,
both high temperature and high values of the adiabatic index
determine a deficit of power on small scales ($k > 0.1\skm$).
The trends found for $\Paa$ are similar to those found by
\cite{mcdonald05,vh06}.

\begin{figure}
  \centering
  \includegraphics[width=1.0\linewidth]{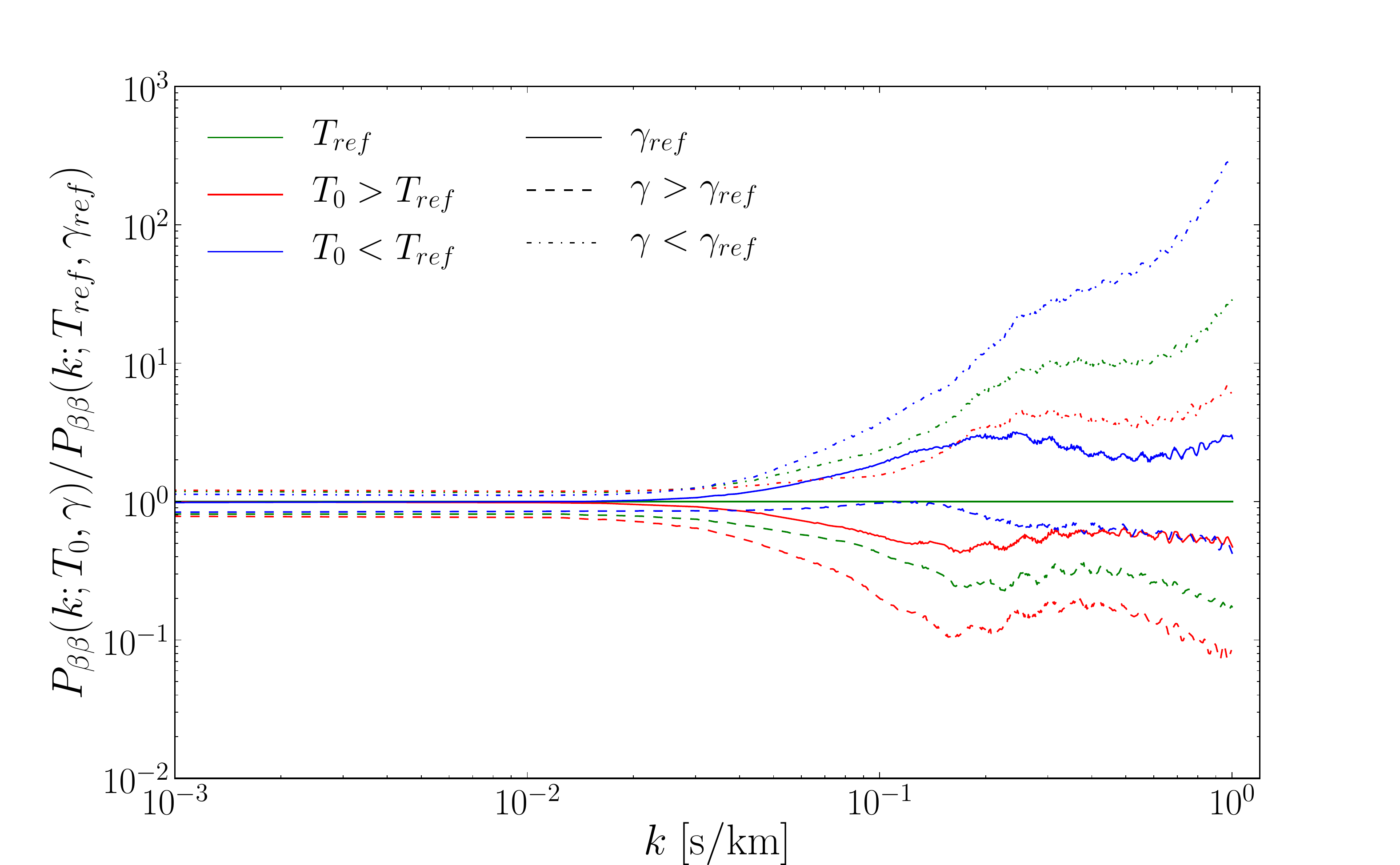}
  \caption{
    \lyb\ power spectrum for different thermal histories (colour
    and line style scheme is the same as that in
    Fig. \ref{Fig:4}). While in general we see similar trends to those
    observed for the $\Paa$ power spectrum, namely hot gas and lower
    $\gamma$ produce deficit of power on small scales the differences
    between the models are much more pronounced at small scales and
    also at large scales where we see clear separation due to
    different values of $\gamma$. This indicates that $\Pbb$ is much
    more sensitive to $\gamma$ than $\Paa$ (see text). 
  }
  \label{Fig:5}
\end{figure}

As with \lya\ power spectrum, similar features can be observed when
looking at the estimate of the \lyb\ power spectrum, shown in
Fig. \ref{Fig:5} (colours and line styles are the same as in
Fig. \ref{Fig:4}). On small scales we can observe the tendency for
both high temperature and high value of $\gamma$ to lower the power
spectrum, while cold IGM and low value of the adiabatic index increase
the power on small scales, when compared to the reference
model. However, there is a noticeable difference between \lya\ and
\lyb\ on large scales ($k < 0.1\skm$): a regime where the difference
between different values of parameters of the equation of state is
more pronounced. In fact, there is a clear separation with the value
of the adiabatic index: high values of $\gamma$ decrease the power,
while small values of $\gamma$ increase the power on large
scales. Whereas the results still remain mostly insensitive (at large
scales) to the changes of IGM temperature. This result can be
interpreted as follows: a small value of the \lyb\ cross section
results in a high sensitivity to large values of density of the
IGM. Even when \lya\ would produce a near saturation for a high enough
density (at a given parameters of equation of state), \lyb\ would
remain in the regime where the transformation between flux and optical
depth remains mostly linear.

We can illustrate this point with a simple toy model. Let us assume
that the estimate of the observed flux ($F$) is the following tracer
of the underlying baryon density field fluctuations ($\delta_b$): 

\be F =
\exp\left[-{\cal A}\left( 1 + \delta_b \right)^{p} \right],
\label{eq:simpleF}
\ee 

with constants ${\cal A}$ and $p$. The amplitude constant $A$ plays a
role of a product of cross section of the line observed and the
average temperature of the IGM, thus ${\cal A} \sim \sigma_i \times T_0^{-0.7}$. On
the other hand, the power-law coefficient $p$ is related to the
adiabatic index $p \sim \gamma$. The toy model uses only local optical
depth, whereas a correct model should then be convolved with a (Voigt)
line profile and corrected for redshift space distortions along the
line of sight. However, for our purposes this simple toy model will
suffice, since convolution will smear the effects described over large
number of scales and redshift space distortions will introduce an
additional cutoff scale in the power spectrum that is dependent mostly on
cosmological parameters.
\begin{figure}
  \centering
  \includegraphics[width=1.0\linewidth]{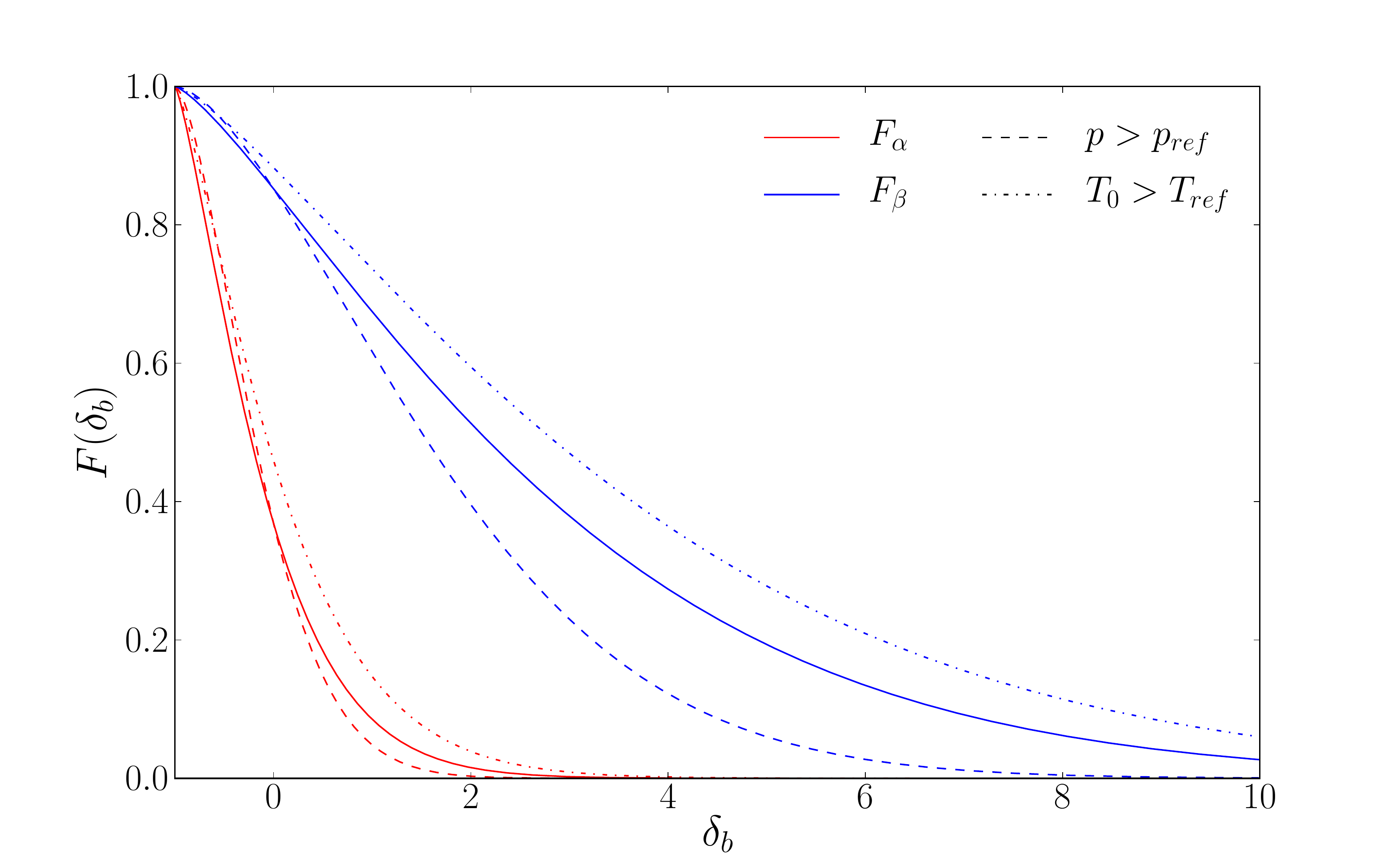}
  \caption{ Dependence of the observed flux on the underlying baryon
    density fluctuation field in a very simple toy model (see
    text). Different colours represent either \lya\ absorption (red)
    or \lyb\ absorption (blue). Different line styles represent
    different values of the model parameters. Full line represents the
    reference values of the parameters ($T_{ref} = 14000$ K and $p_{ref} = 1.3$). 
  }
  \label{Fig:4a}
\end{figure}

In the toy model described, the only difference between \lya\ and
\lyb\ flux estimation will be the value of ${\cal A}$. For any given $T_0$
the ratio of the values of the parameter ${\cal A}$ between \lya\ and
\lyb\ will be the same and equal to the ratio of cross sections of the
absorption lines.
Fig. \ref{Fig:4a} shows the dependence of the observed flux on the
underlying baryon density field ($\delta_b =
\rho_b/\rho_{\mathrm{crit}} - 1$). For given reference values of the
toy model parameters ${\cal A}$ and $p$, the plot shows that \lyb\ flux
covers a wider range of densities than the \lya\ model thus allowing
to be more sensitive to the factor $\left( 1 + \delta_b\right)^p$. We
can also interpret this as if, for a given density level threshold,
the \lya\ flux is largely insensitive to the changes in density, while
for the same threshold \lyb\ is still in a relatively mildly
non-linear regime of sensitivity. The dashed line demonstrates that
changes of the $p$ parameter in the toy model have stronger effects on
the shape of the \lyb\ flux dependence, suggesting that the model is
sensitive to even the small changes of $p$ in \lyb\ flux while not
being sensitive in \lya\ flux. On the other hand, changing the value
of $T_0$ will shift both flux dependencies in a similar fashion,
indicating that both fluxes would be equally sensitive to the change
in $T_0$.  Thus, using the simple arguments above, it is clear that
\lyb\ flux probes higher densities than \lya\ flux due to the fact
that lower cross section allows for higher sensitivity in the same
range of densities as are probed by \lya\ flux. This would explain why
the observed \lyb\ power spectrum (see Fig. \ref{Fig:5}) is more
sensitive to the value of the adiabatic index $\gamma$.

Additionally, in Fig. \ref{Fig:5} an oscillatory pattern can be seen on
very small scales of $k > 0.2\skm$ and is very similar in nature than
pattern noticed in $\Pab$. The results indicate that the
oscillation pattern is more pronounced for hot gas and low
adiabatic index. However, this paper is focused on the $T_0,\gamma$
dependence of the power spectra on large scales ($k < 0.1\skm$) and
we decide not to investigate further the nature of this oscillations. 

\subsection{The cross spectrum dependence on IGM thermal state}
In addition to both the $\Paa$ and $\Pbb$ power spectra we have also
investigated the dependence of $\Pab$ on the thermal history. While the
general statements described so far hold for the cross power spectrum
as well and will be more thoroughly discussed in the following, we
would like to focus on some very peculiar aspects of the cross power
spectrum. 

\begin{figure}
  \centering
  \includegraphics[width=1.0\linewidth]{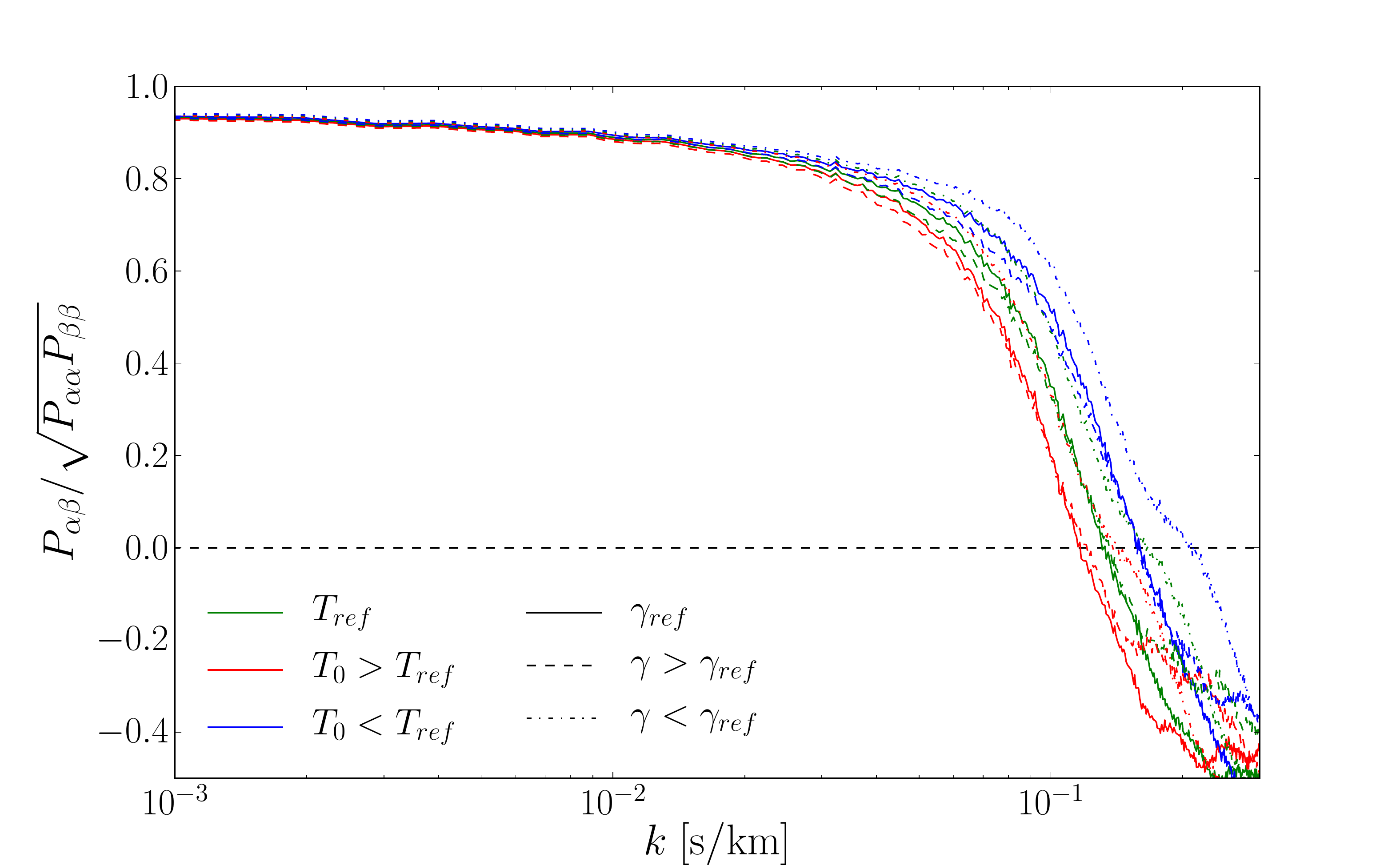}
  \caption{ Normalized cross power spectrum $\Pab$ for different
    thermal histories (colour and line styles as in Figs. \ref{Fig:4}
    and \ref{Fig:5}). Normalized $\Pab$ is very insensitive to
    different thermal histories at largest of scales, even less so
    than $\Paa$. However there is a unique dependence of the
    zero-crossing of the $\Pab$ on the thermal history. For
    intermediate and large values of $\gamma$ the zero-crossing seems
    to depend only on the $T_0$ temperature of the gas. However, this
    is no longer true for small values of the adiabatic index. In a
    similar way as $\Pbb$ is sensitiv to $\gamma$ at large
    scales, it seems that $\Pab$ is sensitive to $T_0$ at somewhat smaller
    scales ($k\sim 0.1\skm$). }
  \label{Fig:6}
\end{figure}

Figure \ref{Fig:6} shows the dependence of the normalized cross power
spectrum for different combinations of the parameters
$(T_0,\gamma)$. An important feature to notice is that the normalized
$\Pab$ is very insensitive to any change in the thermal history on
large scales: this fact has the profound implications that it must
depend on cosmological parameters only. Thus, the normalized $\Pab$
could become a powerful tool to constrain cosmological models
independently of the parameters of the IGM on large scales and should
be thus considered in any future \lya\ forest focused on constraining
cosmological parameters.

Another important feature in Fig. \ref{Fig:6} is the zero-crossing of
$\Pab$. We can see that for intermediate and high values of the
adiabatic index the zero-crossing is determined only by the
temperature of the IGM, while this conclusion is somewhat loosened for
low values of the adiabatic index. Estimating the value of $T_0$
independently of the adiabatic index would be very important for
understanding the properties of the IGM as well as constraining the
parameter space and allowing for better estimates of the cosmological
parameters. However, we caution the reader that the scales of
zero-crossing ($k \sim 0.1\skm \rightarrow r \sim 0.1\Mpch$ are on
the boundary of scales that are confidently retrieved from the current
measurements.

Any future \lya\ surveys with high resolution and signal-to-noise
ratio could use this unique feature of the $\Pab$ to significantly
improve current constraints on $T_0$, since neither \lya\ nor
\lyb\ can constrain $T_0$ alone to high precision.  High resolution
would allow us to push to smaller scales and confidently measure the
scales where the crossing occurs. We stress however that the high
signal-to-noise ratio and a very precise understanding of the
instruments noise is crucial to determine the zero-crossing. This is
because without the precise knowledge of noise properties
zero-crossing becomes crossing of an arbitrary threshold value of the
power spectrum. However, Fig. \ref{Fig:6} shows that $\Pab$ is
sensitive to $T_0$ and insensitive to $\gamma$ only in a small range
around the scales where the cross-power spectrum changes sign. This
high sensitivity of the cross-spectrum to only $T_0$ would be weaker
at other scales.

\subsection{A closer look to the power spectra: quantitative dependence on IGM 
thermal state}
\label{sec:toymodel}
Here, we decide to have a closer look at the large scales.  The scale
dependence of the three power spectra in the range of $0.1 < k < 10\;
\hMpc$ is shown in the following figures (Fig. \ref{Fig:7},
\ref{Fig:8} and \ref{Fig:9}). They are a zoom-ed in versions of the
figures (Fig. \ref{Fig:4}, \ref{Fig:5} and \ref{Fig:6}), focussing on
the larger scales of the retrieved power spectra ($k < 0.1\skm$). All
three power spectra grow constant relative to the reference model,
indicating that on large scales ($k < 0.02\skm$), each of the power
spectrum has the same shape, but a different amplitude, that varies with
thermal history.

On the largest scales of $\Paa$, shown in Fig. \ref{Fig:7}, it is the
colder gas that produces deficit of power, whereas on small scales
colder gas produced an excess of power when compared to the reference
model. This shows that the average temperature of the gas affects the
flux power spectrum differently on different scales, and comes in with
a strong scale dependence that also changes the sign (compared to the
reference value). Comparing Fig. \ref{Fig:4} and \ref{Fig:7} for
\lya\ flux power spectrum, it is possible to write down approximate
scaling relations for the temperature and the adiabatic index.

\begin{figure}
  \centering
  \includegraphics[width=1.0\linewidth]{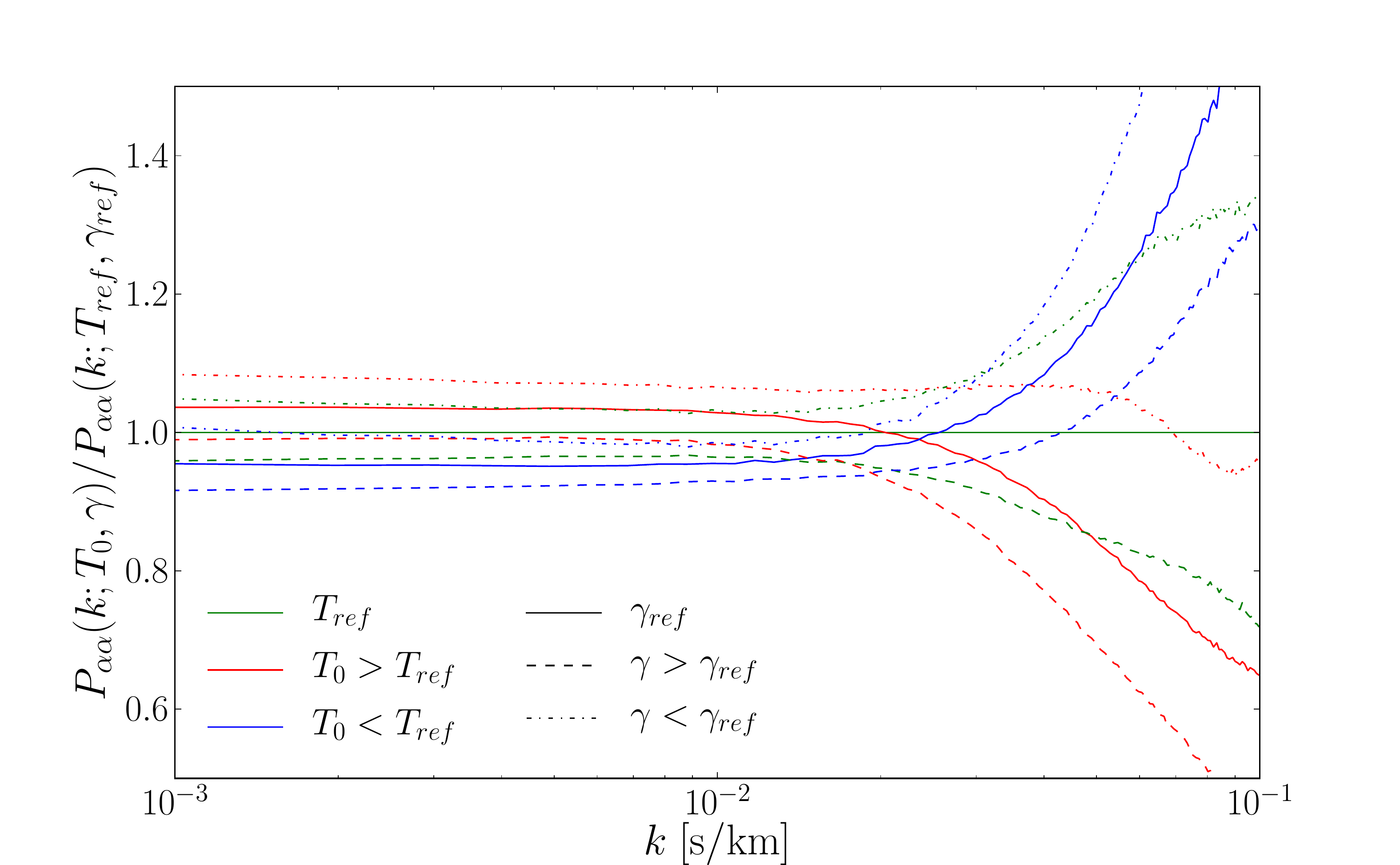}
  \caption{Zoomed-in version of Fig. \ref{Fig:4}: $\Paa$ at large
    scales. Different colours and line styles corresponds to different
    thermal histories: colour corresponding to average gas
    temperature - with red (hot), cold (blue) and green (reference value),
    and line styles corresponding to different values of adiabatic
    index - with dashed (high $\gamma$), dash-dotted (low $\gamma$)
    and full (reference value). Focus is turned more to large scales
    that cover the range of modes further investigated in this
    paper. Unlike smaller scales shown on Fig. \ref{Fig:4} it is
    colder gas (than the REF case) that produces deficit of power on large scales.
  }
  \label{Fig:7}
\end{figure}

The $\Paa$ can be thought of as a function of scale with additional
parameters $T_0$ and $\gamma$. If one would expand $\Paa$ into a Taylor
series over the parameter space around some reference model, the
linear contribution could be written as:
\be
\Paa(k; T_0, \gamma) = \left. \Paa(k) \right|_{\mathrm{ref}} +
\left. \frac{\partial\Paa(k)}{\partial T_0} \right|_{\mathrm{ref}}
\left( T_0 - T_{\mathrm{ref}}\right) +
\left. \frac{\partial\Paa(k)}{\partial \gamma} \right|_{\mathrm{ref}}
\left( \gamma - \gamma_{\mathrm{ref}}\right) + \dots
\label{eq:Paa-taylor}
\ee
The above equation allows us to provide a mathematical context for unveiling
the relation between power spectra and thermal parameters.
Since the
first partial derivatives over parameters depend only on scale and
reference values (but not on the actual values of the parameters), one
can easily interpret the scalings seen in Fig. \ref{Fig:7}. By
keeping the adiabatic index fixed at the reference value, the only
dependence will come from linear Taylor term in the temperature. As
already discussed (Fig. \ref{Fig:7}), at large
scales higher temperature produce excess of power while on smaller
scales it is reverse. The pivot scale around where the behaviour
changes is around $k_p \sim 0.02 \skm$. We can thus approximate the
first derivative term with respect to temperature as:
\be
\left. \frac{\partial\Paa(k)}{\partial T_0} \right|_{\mathrm{ref}}
\approx \left. \Paa(k) \right|_{\mathrm{ref}}\, f(k) \approx \left. \Paa(k) \right|_{\mathrm{ref}}\, A\left( 1 -
\frac{k^2}{k_p^2} \right),
\label{eq:T-1stder}
\ee
where the first term, the power spectrum at the reference value, comes
from the fact that the ratio of power spectra is plotted in
Fig. \ref{Fig:7}. If the amplitude ($A$) of the temperature derivative
term and the pivot scale ($k_p$) are chosen to best reproduce the
simulations we get $A \approx 0.1$ and $k_p \approx 0.02\skm$. We
caution the reader that these values were not produced in a fit of any
sort, nor was the specific scale dependence $f(k)$ chosen to best match the
simulations. Both the parameter values and the function $f(k)$ are
just an illustration in this case. However, they work reasonably well in
determining the general behaviour of the dependence of $\Paa$ on $T_0$
and are very simple in form.

In a similar fashion to the first derivative with respect to
temperature, we can also model the $\gamma$ dependence. Fixing the
temperature of the gas to the reference value, it is clear from
Fig. \ref{Fig:7} that changing the adiabatic index either only
increases or only decreases power on all scales. Inspired by the
result of Eq. \ref{eq:T-1stder}, we can write the first derivative with
respect to the adiabatic index as:
\be
\left. \frac{\partial\Paa(k)}{\partial \gamma} \right|_{\mathrm{ref}}
\approx \left. \Paa(k) \right|_{\mathrm{ref}}\, g(k) \approx
-\left. \Paa(k) \right|_{\mathrm{ref}}\, A\left( 1 +
\frac{k^2}{k_p^2} \right),
\label{eq:g-1stder}
\ee
where the values of $A$ and $k_p$ are the same as in the temperature
dependent term in Eq. \ref{eq:T-1stder}.

From both the toy models and the results shown in Fig. \ref{Fig:7} one
can see that \lya\ flux power spectrum does not distinguish well
between different models of thermal history, especially at large
scales. The difference is at most $\sim 10\%$ and there is a strong
degeneracy between the temperature and the adiabatic index.

Combining Eqns. \ref{eq:T-1stder} and \ref{eq:g-1stder} and
rearranging the terms to collect those with the same power of $k$, one
can show that in the linear model the parameters combination $\Paa$ is
most sensitive to the quantity $T_0 / \gamma$. This further
supports the argument of the strong degeneracy between the two
parameters.

On the other hand, the \lyb\ flux power spectrum is much more
sensitive to thermal history than \lya, as shown in
Fig. \ref{Fig:8}. From the figure it is clear that $\Pbb$ is very
sensitive to the change in adiabatic index on large scales, whereas
there is no clear improvement in the change of $T_0$ compared to
\lya\ flux power spectrum. 

\begin{figure}
  \centering
  \includegraphics[width=1.0\linewidth]{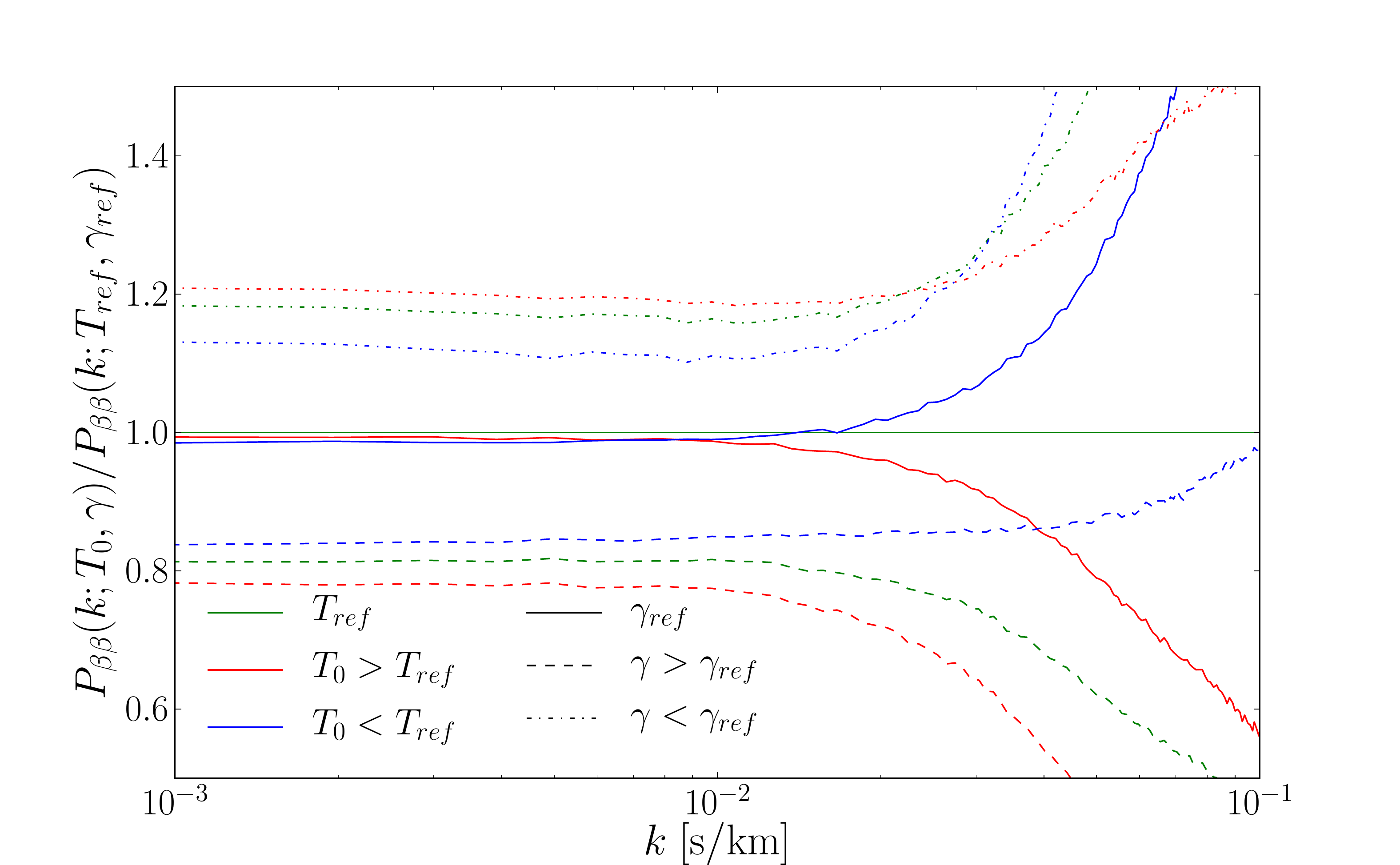}
  \caption{ \lyb\ power spectrum at the large scale (small $k$'s)
    range. The colours and line styles represent the same thermal
    histories as Fig. \ref{Fig:7}. $\Pbb$ is much more sensitive to
    thermal history on large scales than $\Paa$. In fact it is
    strongly sensitive to the choice of the adiabatic index ($\gamma$)
    while not so much to the average temperature of the gas
    ($T_0$). 
    }
  \label{Fig:8}
\end{figure}

A Taylor expansion over the parameters of the equation of state can be
made for $\Pbb$ as well. Similarly to
Eqns. \ref{eq:T-1stder} and \ref{eq:g-1stder} we can write an approximate
model, to underline the general behaviour observed in simulations as:
\be
\Pbb(k; T_0, \gamma) \approx \left. \Pbb(k)
\right|_{\mathrm{ref}}\left[ 1 - A\frac{k^2}{k_p^2}\left(T_0 -
  T_{\mathrm{ref}}\right) - A\left(7 + \frac{k^2}{k_p^2}\right)\left(\gamma -
  \gamma_{\mathrm{ref}}\right)\right],
\label{eq:Pb-1stder}
\ee
where the amplitude $A$ and pivot scale $k_p$ can be chosen to have
the same values as the ones describing $\Paa$. Thus putting $A \approx
0.1$ and $k_p \approx 0.02\skm$ will ensure that the approximation
works reasonably well when it comes to predicting the general
behaviour. However one can quickly see the limitations of the above
equation. Firstly, it is very clear from Fig. \ref{Fig:8} that the
curves representing COLD and HOT model or LG and HG model are not
symmetric around the reference model, thus a simple linear dependence
will not be sufficient to capture this trend. Secondly, the above toy
model (\ref{eq:Pb-1stder})
would indicate that on large scales there is no difference in models
with different average temperature of the IGM. While this is roughly
true for reference model, adding $\gamma$ shows small albeit distinct
separation thus suggesting that either the temperature term has a weak
constant term in addition to the quadratic term in $k$, or that there
is a non-linear term linking $T_0$ and $\gamma$. In fact a cross term
that goes as $\left(\gamma - \gamma_{ref}\right)\,T_0$ might be closest
to the functional form observed in $\Pbb$ power spectrum. Another
important feature to note in Eq. \ref{eq:Pb-1stder} is the high
value of the constant term in $\gamma$ dependence which tries to
explain the observed strong separation of models with different
adiabatic index. 

The fact that \lyb\ is more sensitive to the adiabatic index
($\gamma$) was already pointed out and is most likely linked to the
fact that the nonlinearity of the $F-\tau$ relation enhances the
\lyb\ ability to be sensitive to higher densities. If one would
perform a parameter estimation analysis based on the simulation
results, one would expect a strong constraints on the $\gamma$
coming from \lyb\ flux power spectrum. 

\begin{figure}[H]
  \centering
  \includegraphics[width=1.0\linewidth]{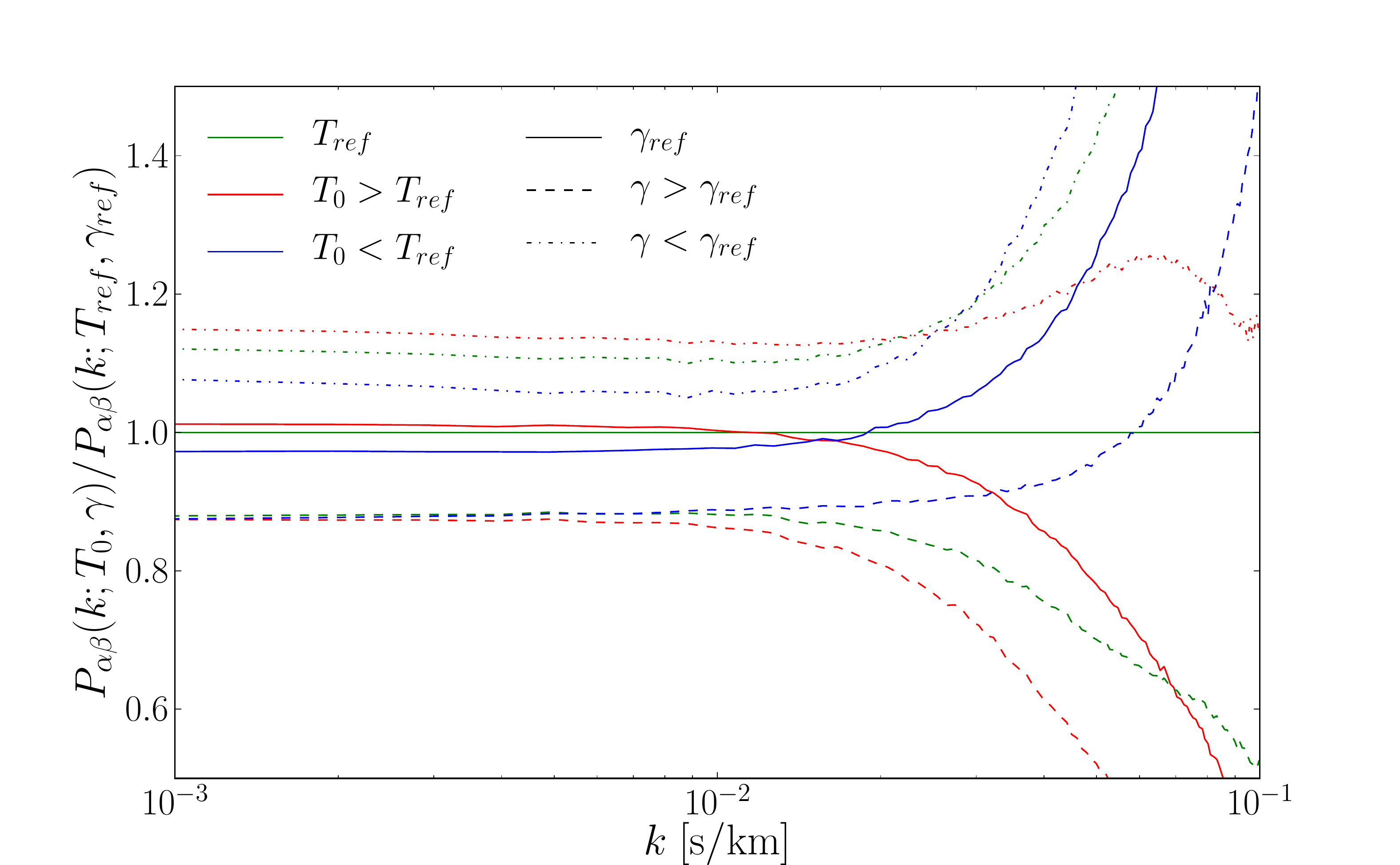}
  \caption{ Plot similar to Figs. \ref{Fig:7} and \ref{Fig:8}
    and shows how (not normalized) $\Pab$ changes with thermal history
    compared to the reference model (REF). The sensitivity to the
    parameters of equation of state is similar to that of $\Pbb$,
    although less pronounced. Also, there seem to be a peculiar
    nonlinear degeneracy between $T_0$ and $\gamma$, making cross
    power nearly insensitive to $T_0$ and high $\gamma$.  }
  \label{Fig:9}
\end{figure}

Once again we stress that these equations
(Eqns. \ref{eq:T-1stder}, \ref{eq:g-1stder} and \ref{eq:Pb-1stder}) are not a
fit to the simulations but merely a toy model, showing the general
trends of changing temperature and the adiabatic index that we see in
simulations. These toy models should not be regarded as tools 
for parameter estimation studies.

The cross power spectrum $\Pab$ shows similar behaviour on large
scales as $\Paa$ and $\Pbb$. As in Fig. \ref{Fig:9}, the cross
power is sensitive to $\gamma$ on large scales, and is between
$\Paa$ and $\Pbb$ in the sensitivity regime, a non-symmetric
dependence on the temperature. The $\Pab$ is sensitive to the
temperature at low values of adiabatic index, whereas at high values
of $\gamma$, it is nearly completely independent of the temperature at
large scales ($k<0.01\skm$). Note that in Fig. \ref{Fig:9} we are
plotting $\Pab$ for different thermal histories divided by the
reference model, unlike in Fig. \ref{Fig:6} where we plotted a
normalized cross power spectrum. While normalized $\Pab$ offers a wide
variety of possible tools to determine the cosmology or the physics of
IGM, in the rest of this paper we focus on using the not normalized $\Pab$
and its information on the large scales. As Fig. \ref{Fig:9} might
indicate, the constraints coming from $\Pab$ on $\gamma$ will be
similar to those of $\Pbb$, while nothing can be concluded about the
constraints on the $T_0$.

\subsection{Adding signal-to-noise ratio and resolution effects}
So far we have shown that all three power spectra have very
interesting and potentially powerful features at both large and small
scales that would help put tighter constraints on the equation of
state parameters and thus allow for better estimation of the
cosmological parameters.

However, the current instruments with medium or low resolution and
reasonable signal-to-noise ratio would undoubtedly set a limit on the
amount of information that could be retrieved. We proceed by
estimating the highest $k$-mode that current surveys would still be
able to reproduce without being contaminated by either imperfect
knowledge of the instrument resolution and/or noise. 

\begin{figure}
  \centering
  \includegraphics[width=1.0\linewidth]{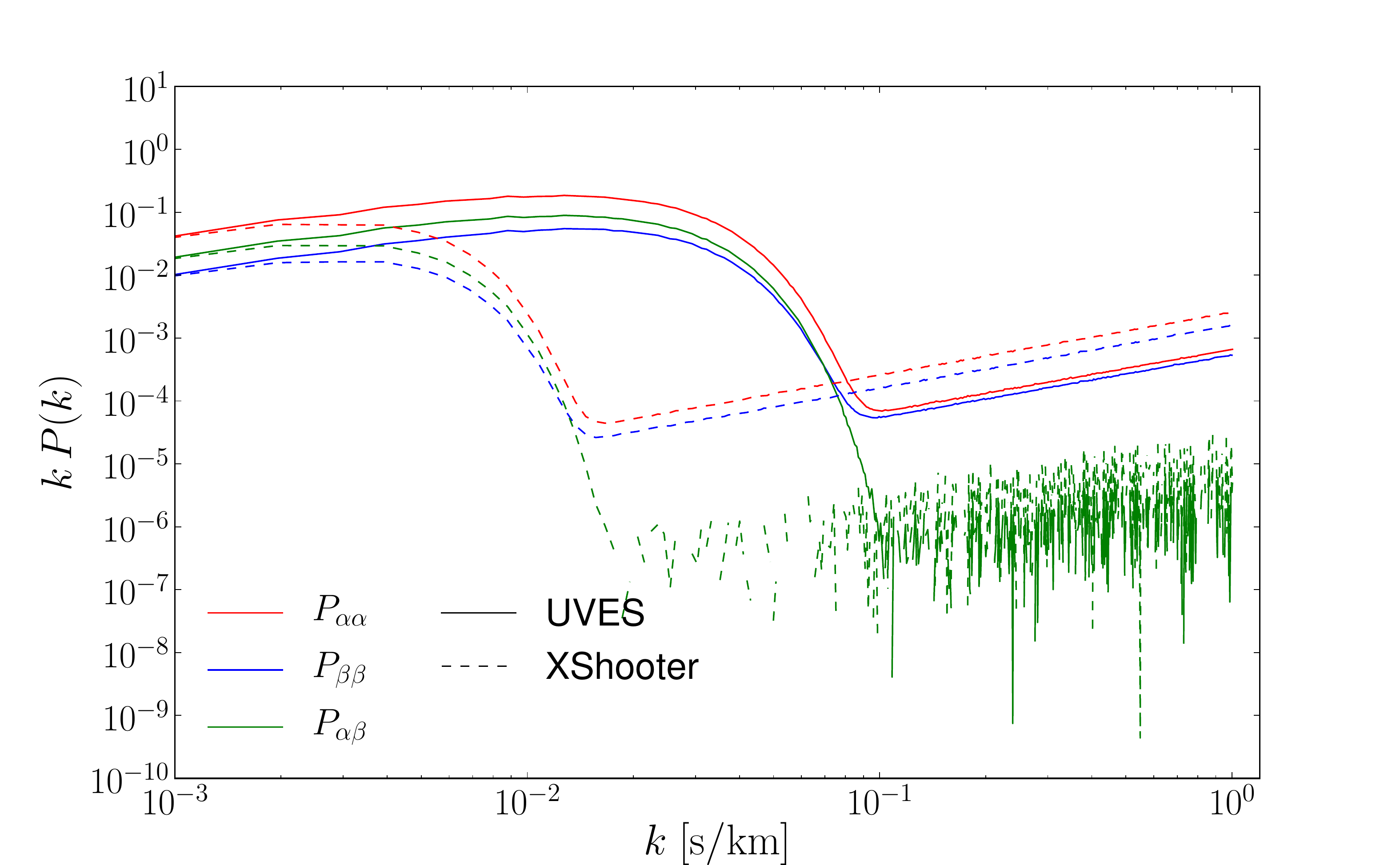}
  \caption{
    $\Paa$ (red),
    $\Pab$ (green) and $\Pbb$ (blue) - at redshift $z=3$ for the
    reference thermal history model (REF) when the flux along each
    sight line has been contaminated with noise and convolved with
    instrumental resolution. The
    instrumental resolution adapted in this plot used Gaussian kernels
    with FWHM of $7\kms$. The added noise was modeled as a flux
    independent part with signal-to-noise ratio of $178$ and flux
    dependent part with signal-to-noise ratio of $89$. These results
    mimicking high resolution and signal to noise data (e.g. UVES) are
    represented in full lines. The flux power spectra for a medium
    resolution and signal-to-noise experiment (e.g. X-Shooter) are
    represented in dashed lines.
  }
  \label{Fig:10}
\end{figure}

Fig. \ref{Fig:10} shows the three power spectra ($\Paa$, $\Pbb$ and
$\Pab$) retrieved from post-processed skewers through simulation
box. The thermal history model used was the reference one. The
post-processing involved adding the specific noise model and
convolving with resolution of real experiments. Each flux LOS has been
convolved with a Gaussian profile of a given FWHM to mimic the effect
of finite resolution of the spectrograph. Furthermore, a randomly
generated Gaussian noise was added to the flux along each line of
sight. The noise was divided into a flux independent part and a flux
dependent part that is present in any higher resolution survey. The
Fig. \ref{Fig:10} shows power spectra (full line) retrieved from a
high resolution spectrograph (such as UVES data as processed in
\cite{k04}) with FWHM $= 7\kms$ and noise model with flux dependent
signal-to-noise ratio of $\left(S/N\right)_F = 89$ and flux
independent $S/N$ ratio of $S/N = 178$. The dashed line shows power
spectra as they would have been retrieved from a medium resolution and
signal-to-noise experiment such as X-Shooter \cite{vernet11}. The
values chosen are conservative for that experiment with FWHM $ =
30\kms$ and $S/N = 10$.

As we can see from the Fig. \ref{Fig:10}, adding realistic noise
properties to the flux from simulations and not accounting for it in
the subsequent analysis of the flux makes it hard to observe
small scale power. Moreover, the addition of resolution effects
introduces a cut-off around the scales of $\sim 0.1\skm$. This is the
scale where the zero-crossing of $\Pab$ occurs as well as most
non-linearities in the $\Paa$ and $\Pbb$. In a realistic experiment,
both effects will be modeled and removed but this plot shows where the
effects will take place if the properties of the noise or resolution
are not known to a high precision or removed correctly.  Thus, for any
future surveys willing to measure the properties of the power spectra
introduced in this paper it will be of the utmost importance to know
and model the noise and resolution of the spectrograph.

In the case where the resolution is known and removed correctly the
limiting factor is the noise of the data. The noise starts to become
important around the scales of $k > 0.1\skm$, and a careful modelling
is required to reproduce the power spectrum at the small scales
because the perturbations giving rise to power on those scales are
much smaller than the noise power. Due to this limiting factor of real
data sets and the fact that current experiments extend only to
around the aforementioned scales we have decided to carefully model
and examine only scales up to $k = 0.1\skm$. Another important factor
in this decision is also the fact that modelling at smaller scales is
increasingly harder due to much more pronounced non-linearities and
also feedback/astrophysical effects might come into play
\cite{mcdonald05,vielschaye}. 

\subsection{Empirical fit to the power spectra}
We have illustrated the general features of the power spectra dependence relative
to some fixed reference model of thermal history in the previous
sections. However, those were toy models barely illustrating the
general parameter dependence and not proper fits to the simulations.

To investigate the parameter dependence of the power spectra we have
adopted a full fitting function, following 
Eq. \ref{eq:Paa-taylor}. The fitting function was based relatively to the reference thermal history model REF (see Table \ref{tb1}). The power
spectrum $P_{i}$ as a function of the mode $k$ with arbitrary
parameters $(T_0,\gamma)$ is thus: \be P_{i}(k;T_0,\gamma) =
\left. P_{i}(k)\right|_{\mathrm{ref}} b_{\mathrm{TH}}^{(i)}(k),
\label{eq:P-fit}
\ee
where the index $i$ runs over all the possible power spectra, i.e. $i
= \{\alpha,\beta,\alpha\beta\}$. The fitting function
$b_{\mathrm{TH}}^{(i)}$ as a function of $k$ was chosen to be a
polynomial of order 3:
\be
b_{\mathrm{TH}}^{(i)}(k) = A^{(i)}\left[1 +
  a_1^{(i)}\left(\frac{k}{k_p}\right) +
  a_2^{(i)}\left(\frac{k}{k_p}\right)^2 +
  a_3^{(i)}\left(\frac{k}{k_p}\right)^3\right],
\label{eq:b-fit}
\ee
where the pivot scale $k_p$ was chosen to be $k_p = 0.05\skm$ and was
fixed in the fitting procedure.
All the coefficients $(A,a_1,a_2,a_3)$ for each of the power
spectra are a linear combination of the thermal history parameters
$(T_0,\gamma)$. This can be summarized as:
\begin{align}
A^{(i)} = A^{(i)}_0 + A^{(i)}_T T_4 + A^{(i)}_\gamma \gamma \notag \\
a^{(i)}_j = a^{(i)}_{j0} + a^{(i)}_{jT} T_4 + a^{(i)}_{j\gamma}
\gamma,
\label{eq:Ak-fit}
\end{align}
where $T_4$ is the average temperature of the gas in the units of
$10^4\K$. The terms of Eqns. \ref{eq:P-fit}, \ref{eq:b-fit} and
\ref{eq:Ak-fit} can be rearranged into ones similar to those of a
Taylor series expansion given in Eq. \ref{eq:Paa-taylor} to see that
in the fitting function adopted in this paper we have extended the
Taylor series expansion to second order in parameters of the
temperature-density relation. 
The above prescription is based on the simple toy models
where we have seen that expansion to linear order in $(T_0,\gamma)$
and quadratic functions in $k$ describe the general behaviour
well. However, this form was not motivated by any physical processes
and there might be a better functional form able to more fully capture the
non-linear behaviour of the power spectra.

\begin{figure}
  \centering
  \subfigure[]{
    \includegraphics[width=0.8\linewidth]{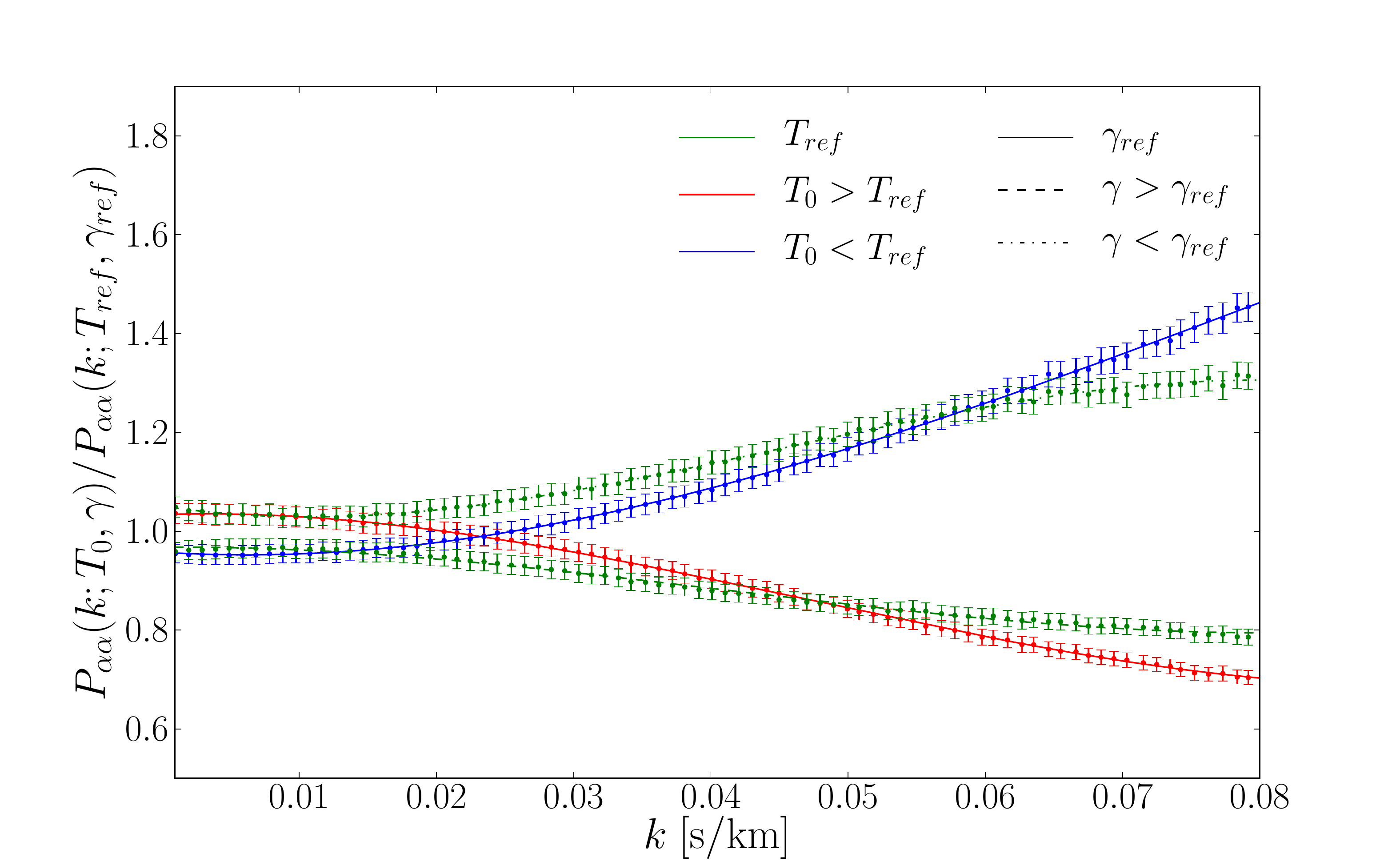}
    \label{Fig:11a}
  }
  \subfigure[]{
    \includegraphics[width=0.8\linewidth]{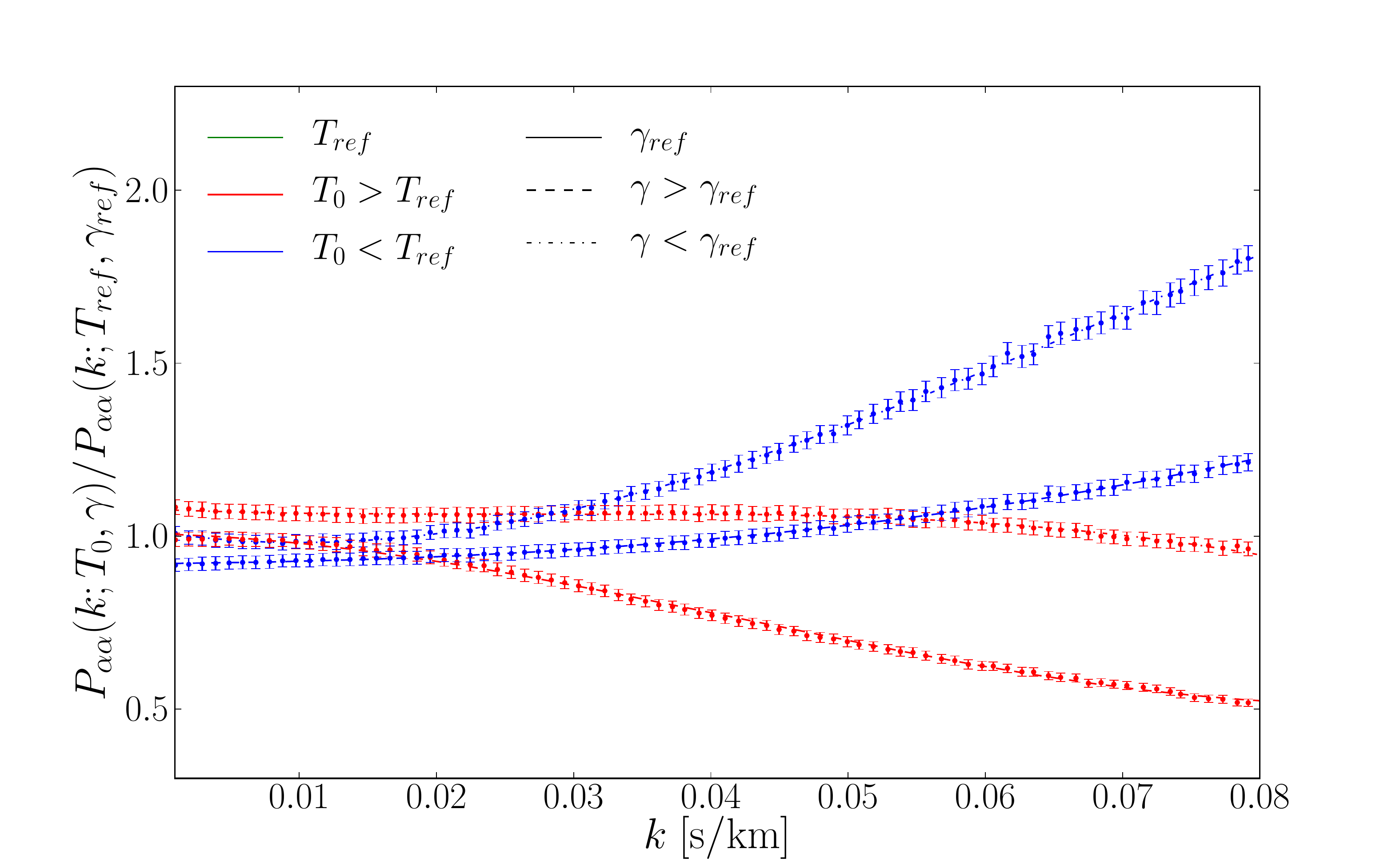}
    \label{Fig:11b}
  }
  \caption{ Fits to the $\Paa$ power spectrum for different thermal
    histories relative to the reference model
    (REF). Fig. \ref{Fig:11a} shows the fits for models: HOT, COLD,
    LG and HG, while Fig. \ref{Fig:11b} shows the results for
    models: HLG, HHG, CLG and CHG. The four parameter model
    fitted is a 3rd order polynomial in $k$, fitted to each mode
    separately. The fitting range extended from the largest of scales
    to $0.08\skm$ (see text). }
  \label{Fig:11}
\end{figure}

A relatively simple form of the fitting function with $12$
parameters for each power spectrum is enough to describe the relative
changes to the power spectra due to the equation-of-state
parameters. In this fit, all the other parameters a power spectrum
might depend on, were fixed (such as cosmological parameters, etc.). 
The range of modes over which the fitting function works
very well and where we have decided to fit for the exact values of the
parameters is for $10^{-3}\skm < k < 0.08 \skm$. 

While in principle we
believe it is possible to use the same fitting function to extend to
larger scales it is not possible to use it on smaller scales. The
nonlinearities induced by both the nonlinear transformation between
the flux and the density field, as well as non-trivial and
highly-nonlinear dependence on the thermal history of the power
spectrum on small scales render this approach impractical and
incorrect on small scales. However, since most of the recent
experiments measure power spectra reliably only up to $k = 0.1\skm$ it
seems sufficient to use that range to show what an improvement an
inclusion of the \lyb\ power spectrum can bring to the estimation of
the equation-of-state parameters. 

Another possibility of an empirical fitting formula with a few more
parameters was proposed by \cite{mcdonald03}. While this method is
capable of fitting the power spectrum directly and is not aimed at
capturing the ratios between different thermal histories it requires a
good estimation of large scales to properly fix all of the model
parameters. However in this work, we have adopted a new, simpler
empirical fit, aimed only at describing the thermal history dependence
of the flux power spectra.

Figure \ref{Fig:11} shows the results of the fitting procedure for
$\Paa$ for all the different thermal histories. It is clear that the
relatively simple fitting function is able to describe the changes of
the power spectrum due to different thermal histories. The fit is
performing well on large scales and starts to get worse on small
scales, as expected since on small scales non-linearities dominate and
are not captured in the fitting function. To account for the
variability of the power spectrum, due to different thermal histories, a
further study is necessary. Also the fit seems to perform better for
different values of average gas temperature $T_0$ while changing
$\gamma$ produces largest discrepancies on small scales, especially
when coupled with high gas temperatures.

The error bars shown in the Fig. \ref{Fig:11} are estimated using
bootstrap method on the extracted LOSs through the simulation box. Only
diagonal elements of the bootstrap matrix were used in the chi-square
estimation of the parameters. Inclusion of the correlations between
the different $k$-bins would only change the errors on the parameters
slightly but would not change their mean value. 

Overall, the estimated parameters given along with the reference
thermal history model (REF) provides a very good description of the
three power spectra ($\Paa$,$\Pbb$,$\Pab$) in the wavenumber range of
$k < 0.08\skm$. The fitted numbers for the parameters are given in \ref{appA}.

\section{Fisher matrix analysis}

Using the empirical fit presented in the previous section we perform
the Fisher matrix analysis on the parameters of the
equation-of-state. The Fisher matrix analysis is widely used to
forecast the precision of an experiment. In our case we will use the
Fisher matrix to forecast the precision with which the equation of
state parameters ($T_0$,$\gamma$) would be extracted from the
simulations using both \lya\ and \lyb\ forests. 

\begin{figure}
  \centering
  \includegraphics[width=1\linewidth]{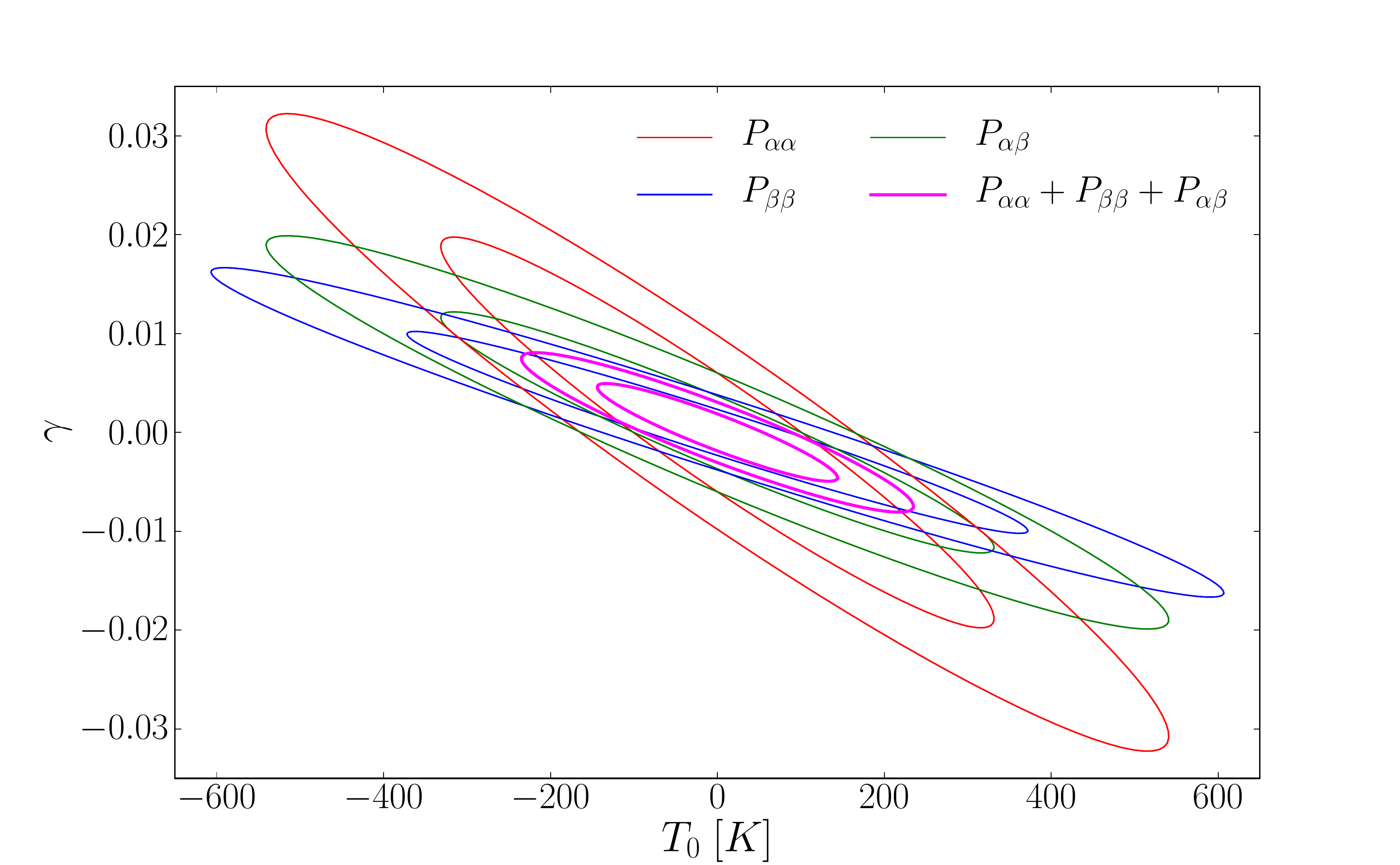}
  \caption{
    Contours of the Fisher matrix analysis of
    recovering the IGM parameters $T_0$ and $\gamma$ using the
    empirical fits presented in this paper. The cosmology is fixed and
    not varied and the only parameters that vary are those describing
    the thermal history. We parameterize the dependence of the three
    power spectra - $\Paa$ (red), $\Pab$ (blue) and $\Pbb$ (green) -
    on the thermal history with 
    just the parameters describing an adiabatic equation of state of
    the IGM. The Gaussian contours obtained from the Fisher matrix
    analysis show us that information on $T_0$ and $\gamma$ is indeed
    constrained in a different and independent way in $\Pbb$ (and $\Pab$)
    than the already existing information using $\Paa$ only. Including
    the data from all three power spectra (magenta) reduces the errorbars on the
    recovered $T_0$ for a factor of $\sim 2-3$ and on $\gamma$ for a
    factor of $\sim 4-5$. This in
    turn would imply a reduction of errorbars on cosmology parameters
    constrained by \lya\ forest ($n_{\rm s}$, $\sigma_8$) by similar factor. 
  }
  \label{Fig:12}
\end{figure}

In this paper we
assume simulations are our experiments in the sense that the former
will predict real data observations of the flux power spectra. Our
analysis makes two assumptions about bootstrap errors on the
simulations: 
we do not take into account the full covariance matrix of the
bootstrap sampling but only the diagonal elements (which we call
bootstrap errors); and we further assume that the bootstrap errors on
the simulations would trace errors on the flux power spectrum from the
real data. At this stage we do not take into account the effects of
real data noise and improper beam corrections because to properly
account for those two effects an analysis should be run on a full mock
data.

Figure \ref{Fig:12} summarizes the main result of this paper and
displays the 1$-\sigma$ and 2$-\sigma$ contours in the plane of the
``equation-of-state'' parameters $T_0$ and $\gamma$. The contours are
centered around the reference mean value points (REF model). Different
colours represent contours obtained from different power spectra or
when combining all the power spectra. From Fig. \ref{Fig:12} it is clear
that there is big ($2-3\times$) improvement to the constraints on $\gamma$ coming
from \lyb\ compared to the ones from \lya\ flux power spectrum. 
It is
important to note is that the \lyb\ flux power spectrum constraints the
parameters space of equation of state in a different way than
\lya\ flux power, and the figure is quantifying this. The constraints
from the real part of the cross power spectrum are similar to those
from \lyb\, yet slightly less stringent. However, they add important
independent information so when combining the information for all
three flux power spectra the constraints on both $T_0$ and $\gamma$
can be tightened considerably (by $4-5\times$ on $\gamma$) 
compared to just using \lya\ flux power.

The Fig. \ref{Fig:12} also shows very clearly the degeneracies between
$T_0$ and $\gamma$ as indicated in Sec. \ref{sec:toymodel}. The
plot nicely shows that $T_0$ and $\gamma$ are strongly (anti-)correlated
in the case of the \lya\ power spectrum, while this degeneracy is
slightly lifted in using \lyb\ power spectrum.

Note that in this paper we have performed all the analysis assuming
equal effective path length for both the \lya\ and \lyb\ optical depth
sight-lines. However, in any real data, the \lyb\ forest path length
would be roughly 20$\%$ the path length of \lya\ forest, which would
result in roughly a factor of $2$ improvement in the errorbars on
$T_0$ and $\gamma$ from \lya\ alone.

\section{Conclusions}
The interest in measuring the IGM thermal state is twofold: on one
side it is of primary importance to constrain the IGM thermal state in
order to lower the impact that such nuisance parameters have in terms
of cosmological derived parameters; on the other side, measuring the
IGM thermal state can shed light on the IGM/galaxy interplay and on
the overall evolution of baryons helping to understand aspects like
feedback and Helium or Hydrogen reionization.  Moving from \lya\ to
\lyb\ forest and relying on different flux statistics (e.g. flux probability distribution function,
bispectrum, curvature of the transmitted flux, etc.) for both the transitions can be a very promising
technique.

It has been first suggested in \cite{lidzlyb} that using \lyb\ power
spectrum one could improve the constraints on the IGM parameters
(e.g. parameters of the $T-\rho_b$ relation). Also following the recent
measurements of the \lyb\ power spectrum \cite{irsic13} we have
performed a more detailed study of \lyb\ forest constraints on the IGM
parameters.

Using a suite of high resolution hydrodynamic simulations, we have
investigated how the \lya\ and \lyb\ flux power spectra change in the
presence of different thermal histories. We have also included the
(real part of the) cross power spectrum in the calculations.

The main findings of this paper can be summarized as follows.
\begin{itemize}
\item[(i)] For a given thermal history all three power spectra $\Paa$,
  $\Pbb$ and $\Pab$ show constant bias among each other on large
  scales ($k < 0.01\skm$). If $\Pbb$ (and $\Pab$) are biased tracers
  of $\Paa$ (and thus of the underlying matter distribution) then the
  bias on large scales would
  depend only on the different sensitivity of the $\Paa$ and $\Pbb$ to
  the IGM thermal parameters.
\item[(ii)] The real part of the cross power spectrum $\Pab$ displays
  a distinct oscillation pattern at small scales ($k>0.01\skm$). The
  oscillation pattern is such that $\Pab$ crosses zero and becomes
  negative at a specific range of scales.
\item[(iii)] The normalized cross power $\Pab/\sqrt{\Paa \Pbb}$ is
  insensitive to thermal history on large scales
  ($k<0.01\skm$). Furthermore, the crossing of zero is sensitive only to
  $T_0$ at high and intermediate values of $\gamma$. However, this
  specific property is only true on a small scale range around the (first)
  crossing of zero ($k \sim 0.1\skm$).
\item[(iv)] The dependence of $\Pbb$ on thermal history is much more
  pronounced than for $\Paa$. As explained in the paper we believe
  this is because the \lyb\ absorption probes denser regions due to
  smaller optical depth of the \lyb\ absorption cross-section. This is
  most simply seen in the fact that $\Pbb$ is very sensitive to
  different values $\gamma$, suggesting that the values of the
  overdensities probed are typically larger than the corresponding
  ones probed in $\Paa$.
\item[(v)] Using the Fisher matrix formalism we show that the $\Pbb$
  alone is more sensitive to $\gamma$ than $\Paa$, putting $2-3$
   times more stringent constraints on that IGM parameter. Combining
  the information from all three power spectra improves the estimation
  compared to using only $\Paa$ even further to achieve $4-5$ times  tighter
  constraints on the slope of the $T-\rho_b$ relation and $2-3$ times
  tighter constraints on the mean average temperature.
\end{itemize}

While in this paper we show that $\Pbb$ is more sensitive to
thermal IGM parameters, particularly the slope of the
temperature-density relation, our analysis has still to be intended as
preliminary. For example, in our analysis we have fixed the cosmology
to the cosmic microwave background values, so it is not clear by which
amount the $\Pbb$ could improve the constraints on the cosmological
parameters. This is of course of much interest and further analysis
should be conducted to investigate the joint parameter
space. Moreover, our empirical fit is parametrized only by parameters
of the $T-\rho_b$ relation and thus does not take into account the
damping due to Jeans filtering which is a more complicated
(integrated) function of the thermal history of the universe. Finally,
our assumption is that the $T-\rho_b$ relation can be described by a
simple power-law relation and this can break down around HeliumII
reionization (e.g. \cite{compostella,mcquinn}). It will be also
important to test the full procedure on top of more refined mock
quasar spectra that incorporate all the relevant instrumental effects
in a similar way as it is presently done in the BOSS collaboration
(see e.g. \cite{font12}).

An interesting point that we have not further investigated in this paper is
the effect of the scatter in the $T-\rho_b$ relationship on the final
result. Much like the flux power spectrum, temperature-density
relation is another important analysis tool that can be used to
characterize the physics of the IGM. In this work, however, we have
only used the $T-\rho_b$ scatter plot in a regime where the power-law
approximation is valid, and the scatter is small. The values of the
parameters extracted in this way are then used as a parameterization of
the flux power spectrum (for both \lya\ and \lyb). One might worry
that since \lyb\ flux power spectrum is more sensitive to higher
density regions, it would be more affected by the scatter in the $T-\rho_b$
relationship. However, the analysis presented in this paper does not aim at
constraining the temperature-density relation, but only a possible
parameterization of it. The results of the fitting functions and
Fisher matrix analysis are in that sense completely independent of
scatter. Nevertheless, one might construct a different approximation
of the $T-\rho_b$ relationship, going beyond the simple power law
approximation (e.g. stochastic or asymmetric scatter or density
dependent scatter). In such a case one could parameterize the relation
with more than just $T_0$ and $\gamma$ and possibly, by using both
\lya\ and \lyb\ flux power spectrum, it might put some
constraints of the scatter of the underlying IGM $T-\rho_b$
relation. The scatter would then act as an additional parameter (or
parameters if the scatter model was complex) that could be used to
expand the flux power spectrum around the reference model. It is
definitely an interesting possibility that would allow us to probe
even more complex approximations of the $T-\rho_b$ relation, using the
\lya\ and \lyb\ power spectrum.

Also we have not investigated  the special properties of the
cross power spectrum $\Pab$ at relatively small scales. The $\Pab$ offers a unique probe
of the temperature of the IGM but the relation of the zero crossing
needs to be investigated further, for example by varying cosmological
parameters as well. More work is thus required to investigate the issues presented
here and to fully exploit extremely promising potential of the \lyb\ forest in measuring
the properties of the IGM and in constraining cosmological parameters.

\section*{Acknowledgments}
We thank Jamie Bolton for having provided the thermal histories
routine used in the present paper and for useful discussions.  MV is
supported by the ERC Starting Grant ``CosmoIGM'' and by the IS PD51 INDARK
INFN grant.

\vspace{0.5cm}
\section*{Appendix: The values of the fitted parameters}
\label{appA}

\renewcommand{\theequation}{A\arabic{equation}}
\setcounter{equation}{0}
The parameter values obtained from a fit to $\Paa$ power spectrum are:
\begin{align}
A^{(\alpha)} &= 1.07478^{+0.02229}_{-0.02229} +
0.07864^{+0.00769}_{-0.00769} \times T_4 -
0.14009^{+0.01383}_{-0.01383} \times \gamma \notag \\
a_1^{(\alpha)} &= -0.11272^{+0.02319}_{-0.02319} +
0.01271^{+0.00816}_{-0.00816} \times T_4 +
0.06917^{+0.01448}_{-0.01448} \times \gamma \notag \\
a_2^{(\alpha)} &= 0.10633^{+0.00679}_{-0.00679} -
0.02387^{+0.00237}_{-0.00237} \times T_4 -
0.05238^{+0.00423}_{-0.00423} \times \gamma \notag \\
a_3^{(\alpha)} &= -0.00704^{+0.00056}_{-0.00056} +
0.00121^{+0.00019}_{-0.00019} \times T_4 +
0.00387^{+0.00035}_{-0.00035} \times \gamma,
\end{align}
where the parameter errors quoted are the 1-$\sigma$ errorbars. We can
see that all parameters are estimated with relatively high
significance. However, the temperature dependence of the $a_1$
parameter seems to be poorly estimated, indicating that the $a_1$
does not depend on the $T_0$ or that its dependence on $T_0$ does not
change the final chi-square much.

With the same fitting function we also fitted for $\Pbb$ power
spectrum, obtaining the following parameter values:
\begin{align}
A^{(\beta)} &= 1.54886^{+0.02436}_{-0.02436} +
0.03658^{+0.00837}_{-0.00837} \times T_4 -
0.44332^{+0.01539}_{-0.01539} \times \gamma \notag \\
a_1^{(\beta)} &= -0.117523^{+0.02669}_{-0.02669} -
0.00892^{+0.00932}_{-0.00932} \times T_4 +
0.09610^{+0.01672}_{-0.01672} \times \gamma \notag \\
a_2^{(\beta)} &= 0.12781^{+0.00811}_{-0.00811} -
0.01818^{+0.00280}_{-0.00280} \times T_4 -
0.07434^{+0.00508}_{-0.00508} \times \gamma \notag \\
a_3^{(\beta)} &= -0.00707^{+0.00045}_{-0.00045} +
0.00045^{+0.00023}_{-0.00023} \times T_4 +
0.00471^{+0.00043}_{-0.00043} \times \gamma,
\end{align}
where the errors quoted are again the 1-$\sigma$ errors on the
parameters. The parameters are also slightly correlated. However the
errorbars are much smaller than the bootstrap errors on the power
spectrum bins, and thus any estimation of the ($T_0,\gamma$)
parameters using the above empirical fit would be dominated by the
errorbars on the power spectrum bins.

The same procedure as for $\Paa$ and $\Pbb$ power spectra was also
carried out for the cross power spectrum $\Pab$. The estimated
parameter values are:
\begin{align}
A^{(\alpha\beta)} &= 1.98392^{+0.02355}_{-0.02355} -
0.01157^{+0.00813}_{-0.00813} \times T_4 -
0.71124^{+0.01513}_{-0.01513} \times \gamma \notag \\
a_1^{(\alpha\beta)} &= -0.16986^{+0.02473}_{-0.02473} +
0.00536^{+0.00871}_{-0.00871} \times T_4 +
0.11914^{+0.01559}_{-0.01559} \times \gamma \notag \\
a_2^{(\alpha\beta)} &= 0.15282^{+0.00724}_{-0.00724} -
0.02449^{+0.00251}_{-0.00251} \times T_4 -
0.08592^{+0.00456}_{-0.00456} \times \gamma \notag \\
a_3^{(\alpha\beta)} &= -0.00943^{+0.00060}_{-0.00060} +
0.00144^{+0.00021}_{-0.00021} \times T_4 +
0.00537^{+0.00038}_{-0.00038} \times \gamma,
\end{align}
where the errorbars are the 1-$\sigma$ errors. In the parameter estimation
for both $\Pbb$ and $\Pab$ we can see that the temperature dependence
of $a_1$ is not well constrained at all, giving further indication
that in the region fitted by the empirical function
(Eq. \ref{eq:b-fit}) the parameter in front of the linear $k$-term does not depend on the temperature.

\bibliographystyle{JHEP}
\bibliography{Bibliofile}

\end{document}